%% file: tRecX-CPiP.tex
\documentclass{elsarticle}
\pdfoutput=1
\usepackage{lineno,hyperref}
\modulolinenumbers[5]

\usepackage{color}
\newcommand{\change}[2]{{}{#2}}

\journal{Journal of \LaTeX\ Templates}

\usepackage{array}
\usepackage{amsbsy}
\usepackage{amsmath}
\usepackage{amssymb}
\usepackage{makeidx}
\usepackage{graphicx}
\usepackage{listings}
\usepackage{tikz}
\usetikzlibrary{arrows.meta}
\usetikzlibrary{shapes,arrows}

\include{my_commands}

\renewcommand{\myoper}[1]{{#1}}

\newcommand{\upp}[1]{^{#1}}
\newcommand{\bas}{\underline{b}}
\newcommand{\Bas}{\underline{B}}
\newcommand{\Grid}{\overline{G}}
\newcommand{\Coef}{\overline{C}}
\newcommand{\mvals}[1]{b^{J_{#1}}_{\al_{#1}j_{#1}}}
\newcommand{\mgrid}[1]{\Grid_{A_{#1}}^{J_{#1}}}
\newcommand{\aleg}[2]{P^{|#1|}_{#2}}
\newcommand{\au}{a.u.\ }

\newcommand{\lcode}[1]{{\tt\detokenize{#1}}}

\begin{document}

\begin{frontmatter}



%
%

\title{tRecX --- an environment for solving time-dependent Schr\"odinger-like problems}
 
\author{Armin Scrinzi\corref{author}}
\address{Ludwig Maximilian University, Theresienstrasse 37, 80333 Munich, Germany}
\cortext[author] {\textit{E-mail address:} Armin.Scrinzi@lmu.de}

\begin{abstract}
tRecX is a C++ code for solving generalized inhomogeneous time-dependent Schr\"odinger-type equations 
$id\Psi/dt  = H[t,\Psi] + \Phi$ in arbitrary dimensions and in a variety of coordinate systems. 
The operator $H[t,\Psi]$ may have simple non-linearities, as in Gross-Pitaevskii and Hartree(-Fock) problems.
Primary application of tRecX has been non-perturbative strong-field single and double
photo-electron emission in atomic and molecular physics. The code is designed for large-scale {\it ab initio} calculations, 
for exploring models, and for advanced teaching in computational physics. Distinctive numerical methods are 
the time-dependent surface flux method for the computation of single and double emission spectra and 
exterior complex scaling for absorption. Wave functions 
and operators are handled by tree-structures with the systematic use of recursion on the coarse-grain level. 
Numerical, analytic, and grid-based discretizations can be combined and are treated on the same abstract level. 
Operators are specified in the input using a script language including symbolic algebra. 
User-friendly in- and output, error safety, and documentation are
integrated by design.
\end{abstract}

\begin{keyword}
\texttt Schr\"odinger solver  \sep strong field physics \sep attosecond physics \sep recursive structure
\end{keyword}

\end{frontmatter}

\begin{small}
\noindent
{\bf PROGRAM SUMMARY}
\\
{\em Program title:} tRecX --- time-dependent Recursive indeXing (tRecX=tSurff+irECS)
\\
{\em CPC Library link to program files:} (to be added by Technical Editor)
\\
{\em Developer's repository link:} https://gitlab.physik.uni-muenchen.de/AG-Scrinzi/tRecX 
\\
{\em Code Ocean capsule:} (to be added by Technical Editor)
\\
{\em Licensing provisions:} GNU General Public License 2
\\
{\em Programming language:} C++
\\
{\em Nature of problem}: tRecX is a general solver for time-dependent Schr\"odinger-like problems, 
with applications mostly in strong field and attosecond physics. There are no technical restrictions on 
the spatial dimension of the problem with up to 6 spatial dimensions realized in the strong-field double 
ionization of Helium. A selection of coordinate systems is available and any Hamiltonian involving up to second derivatives 
and arbitrary up to three dimensional potentials can be defined on input by simple scripts.
\\
{\em Solution method:} The method of lines is used with spatial discretization by a flexible combination of one dimensional basis sets, 
DVR representations, discrete vectors, expansions into higher-dimensional eigenfunctions of user-defined operators and multi-center basis sets. 
Photo-emission spectra are calculated using the time-dependent surface flux method (tSurff) in combination with infinite 
range exterior complex scaling (irECS) for absorption. The code is object oriented and makes extensive use of tree-structures and recursive algorithms. 
Parallelization is by MPI. Code design and performance allow use in production as well as for graduate level training.
\end{small}

\tableofcontents

\section{Introduction}

The tRecX code package is designed to be a high-performance, yet flexible and robust code with good maintainability and usability
for Schr\"odinger-like time-dependent problems. It is in use for computing the interaction of atomic and molecular systems
in non-perturbatively strong laser fields. It implements a range of techniques such as irECS (infinite-range exterior complex
scaling \cite{Scrinzi2010}), tSurff (the time-dependent surface flux method \cite{Tao2012,scrinzi12:tsurff}), general and mixed
gauges \cite{majety15:gauge}, and the FE-DVR method for complex scaling \cite{rescigno00:fem-dvr,weinmuller17:dvr}. 
The hybrid anti-symmetrized Coupled Channels
method (haCC \cite{majety15:hacc}) is going to be made publicly available with the next release. The code has been developed for and 
applied to solving several problems in strong field physics. The most outstanding applications of tRecX are the computation of fully-differential 
double electron emission spectra of the Helium atom \cite{zielinski16:doubleionization,zhu20:doubleionization} at laser wave length 
from 10 to 800 nm, including also elliptically polarized
fields \cite{majety17:heElliptic}, as well as strong field ionization rates and photo-emission spectra for di- and tri-atomic linear molecules \cite{majety15:dynamicexchange,majety15:static,majety17:co2spectra} 
with arbitrary alignment between the direction of laser polarization and the molecular axis.

During the development of the code a conscious effort has been and still is being made to adhere to good programming 
practice for ensuring re-usability and maintainability. The object-oriented C++ code
systematically uses abstract and template classes for ensuring uniform and transparent code structure. For 
easier accessibility by physicists these classes reflect concepts that are familiar in physics such as the linear and more specifically 
Hilbert space, operators that are usually but not necessarily linear maps, and wave functions.
Discretization of the wave function is in terms of an abstract basis set class, whose specific implementation
covers the whole range from discrete sets of vectors, over grids, finite-elements, standard basis sets 
such as spherical harmonics, all the way to expansions in terms of eigenfunctions of a user-defined operator.
These can be combined in a tree-structured hierarchy that admits building correlated (non-product) bases from one-dimensional factors.
For performance, numerical libraries such as Lapack \cite{lapack}, Eigen \cite{eigen}, or FFTW \cite{fftw} are used on the low level. Parallelization
is through MPI with some degree of automatic load-balancing based on self-measurement of the code. 

The development of tRecX was initially motivated by several simultaneous PhD projects all related to the time-dependent
Schr\"odinger equation (TDSE), but varying in dimension from 1 to 6 with different coordinate systems and discretization 
strategies. Using math-type strings for input of discretizations and operators allowed covering all these projects
within the same framework, reducing supervision overhead, code redundancy, and  programming errors. 
Also, non-trivial model Hamiltonians can be implemented quickly with little compromise in computational performance. 
This includes, for example, Floquet calculations or simple many-body systems. 

For the use with research students and for graduate level teaching, but also for productivity in research, error-safety and usability are important design goals, as is adhering to good programming practice. Due to its origin in multiple research projects the code does contain important sections that do not conform to such best practice, but there is an ongoing effort to re-implement those sections with modern standards. Documentation relies on code-readability and Doxygen \cite{doxygen} inline documentation. Input is exclusively through a dedicated class, which only allows documented input and machine-generates up-to-date help. Input
can be as numbers or algebraic expressions with standard mathematical functions and combining SI, cgs(ESU), or atomic units (a.u.). 

\subsection{Purpose and scope of this paper}

We give an overview of typical uses of tRecX that do not require any extensions to the code.
In addition, the code's potential is made clear and possible advanced use with or without code extensions 
is indicated. Far from attempting complete documentation in this place, we expose the mathematical background, logical 
structure, and the principles for mapping equations into the code. Some room is given to describing code structure 
and selected classes. This information, apart from being useful in its own right, is meant to illustrate design philosophy and
principles, which we consider as a defining constituent of the tRecX project.

The aim is to provide answers and/or useful information regarding the following questions:
\begin{itemize}
\item What has been done and what is typically done using tRecX?
\item Is there a possibility of using or adapting tRecX for my problem?
\item What are the most important methods in tRecX? Which of them are specific for tRecX?
\item What is the code structure? How could I extend this for a new use?
\end{itemize}
We do not discuss here specific algorithms or numerical methods in greater detail, giving the relevant references instead.

In the following, we first list examples of applications and discuss the corresponding inputs, before introducing the main methods
used in the code. Finally, code concept and structure are illustrated at the example of its main classes. 
Independent reading of the sections is aided by ample cross-referencing and minor redundancies 
between the sections. No effort is
made to provide a complete manual for the code here or elsewhere. Rather, 
all examples shown and further introductory and advanced examples are provided as tutorials with the code. 
This together with code readability and generous Doxygen annotation is intended to serve as a source of full documentation.

\section{Application examples}
\label{sec:applications}
The code source resides on a git repository \cite{tRecXgit} from where up-to-date information on file structure and compilation 
should be drawn. We only single out 
the subdirectory \lcode{tutorial} that contains input files for a range
of applications, named  \lcode{00HarmonicOsc1.inp}, \lcode{01HarmonicOsc2.inp}, etc.,
where \lcode{tutorial/00} through \lcode{11} systematically introduce the most important input features and 
code functionalities. 

\subsection{A single-electron atom in a strong laser field}
\label{sec:exampleSingle}
We choose the single-electron system for introducing the general 
characteristics of strong field physics problems,
the discretization strategy, and the form of operators in tRecX. 
Complete input at slightly different parameters is given in \lcode{tutorial/11shortPulseIR}.

The single-electron time-dependent Schr\"odinger equation (TDSE) 
in strong fields is, in atomic units (\au)
\begin{equation}\label{eq:tdse1}
i\ddt \Psi(\vr,t) = [-\frac12 \Delta +iA_z(t)\ddz + V(|\vr|)]\Psi(\vr,t).
\end{equation}
This describes an electron bound by a rotationally symmetric potential,
where the laser field is linearly polarized in $z$-direction  $\vEf(t)=(0,0,E_z(t)$ and 
the interaction is written in dipole approximation and velocity gauge with the vector potential
\begin{equation}
\vA(t)=\int_{-\infty}^t d\tau \vEf(\tau).
\label{eq:potA}
\end{equation}
At optical or near-infrared wave length the duration of one field oscillation is on the scale of 100 \au and pulse durations reach 1000's of \au. In ionization, a wide range of momenta appears and the wave function expands to very large size during the pulse. This requires reliable absorption at the simulation box boundaries, if exceeding simulation sizes are to be avoided. In tRecX, the standard method for absorption 
is irECS (Sec.~\ref{sec:tsurff}), which allows to work with box sizes
of only a few 10's of \au, although the underlying problem expands to 1000's of \au.

Technically, "strong field" also means that rotational symmetry is strongly broken. 
Still, the use of polar coordinates and an expansion into spherical harmonics $Y^m_l$ often remains convenient
and efficient.  The ansatz is
\begin{equation}
\Psi(\vr,t)=\sum_{m=-M}^{M}\sum_{l=|m|}^{L} Y\upp{m}_{l}(\varphi,\th)\frac1r\chi^{ml}(r,t).
\end{equation}
In linear polarization the $m$-quantum number is conserved and the problem is effectively two-dimensional. 
The radial functions $\chi^{ml}$ need to support a broad range of momenta, which suggests the use of 
higher order grid methods with sufficient density of points. The standard choice in tRecX is a finite-element discrete-variable method (FE-DVR) with K=10-20 collocation points per element. The density of points is problem-dependent, typical average densities are 2 points per atomic unit. Such an expansion is written as
\begin{equation}
\chi^{ml}(r,t) = \sum_{n=0}^{N-1} \sum_{k=0}^{K-1} b^{n}_{k}(r) C\upp{mlnk}(t).
\end{equation}
We remark here that indices of coefficients and partial wave functions are generally written as superscripts, while basis functions
are labeled by a subscript that counts the basis, and a superscript, that designates the set of basis functions to which the 
individual function belongs. That principle is loosely adhered to throughout the paper and broken occasionally for aesthetic
reasons.

FE-DVR can be considered as a local basis set discretization with Lagrange polynomials as the basis functions on intervals 
$[r^n,r^{n+1}]$
\begin{equation}
b\upp{n}_{k}(r)=L_k\left(\frac{r-r^n}{r^{n+1}-r^n}\right)\text{ for }r\in[r^n,r^{n+1}], 
\qquad 
L_k(y)=\prod_{{j=0,j\neq k}}^{K-1}\frac{y-y_j}{y_k-y_j}.
\end{equation}
The $y_j$ are the quadrature points for a Lobatto quadrature rule on the interval $[0,1]$.
It is sufficient to ensure continuity at the $r_n$, which amounts to a linear constraint on the
expansion coefficients of the form $C^{ml,n-1,K-1}=C^{ml,n,0}$. 

Using polar coordinates for $\vr$, the full expansion can be written as a hierarchy of sums
\begin{equation}\label{eq:expansion}
\Psi(\varphi,\cos\th,r;t)=
\sum_{m=-M}^{M} e^{im\varphi}\sum_{l=|m|}^{L} \aleg{m}{l}(\cos\th)\sum_{n=0}^{N-1} \sum_{k=0}^{K-1} 
\frac{b\upp{n}_{k}(r)}{r} C^{mlnk}(t),
\end{equation}
where $\aleg{m}{l}$ are properly normalized associated Legendre functions. For the computation
of matrix elements all operators involved can be written as (short sums of) tensor products, for example
\begin{equation}\label{eq:laplacePolar}
-\De= 
-\one\otimes\one\otimes\frac1r \ddr^2 r 
-\left(\one\otimes\frac{\pde}{\pde\cos\th}{\sin^2\th}\frac{\pde}{\cos\th}
+\frac{1}{\sin^2\th}\otimes\pde^2_\varphi\right)\otimes \frac{1}{r^2}.
\end{equation}
In this form matrix elements only involve one-dimensional
integrations, which, in FE-DVR, are performed using the underlying Lobatto quadrature scheme. 
We denote quadrature schemes by pairs of nodes and weights, in present example as $(r_j,w_j)$.
For correct results 
in FE-DVR one must use the explicitly symmetric form of any operator involving derivatives. 
For example, one writes
\begin{eqnarray}
\lefteqn{\int_{r_n}^{r_{n_1}} r^2 dr \frac1r b\upp{n}_{k}(r) [-\frac{1}{r}\ddr^2 r \frac1r b\upp{n}_{l}(r)]\to}
\nonumber\\
&&\int_{r_n}^{r_{n_1}} dr [\ddr b\upp{n}_{k}(r) ][\ddr  b\upp{n}_{l}(r)]
=\sum_{j=0}^{K-1} w_j [\ddr b\upp{n}_{k}(r_j)] [\ddr b\upp{n}_{l}(r_j)],
\label{eq:symmetricDerivative}
\end{eqnarray}
and similarly for the other coordinates. Note that in this example the Lobatto quadrature rule gives the exact integral.

For product bases, matrices corresponding to tensor products are tensor products of matrices.
Typical bases in tRecX are {\em not} tensor products, but rather show tree-like interdependence (Sec.~\ref{sec:discretization}).
Still, matrix-vector multiplications can be performed with essentially the same operations count as for strict tensor products  
(cf.~Sec.~\ref{sec:quadratures}). 
In the given case, rotational symmetry of the potential and dipole selection rules reduces operator matrices to simple 
block-tridiagonal matrices and there is no computational advantage in exploiting the tensor-product form.

The negative Laplacian Eq.~(\ref{eq:laplacePolar}) can be specified on input by the string 
\begin{lstlisting}
<1><1><d_1_d>+<1><d_(1-Q*Q)_d><1/(Q*Q)>...
           ...+<d_1_d><1/(1-Q*Q)><1/(Q*Q)>.
\end{lstlisting}
The  pairs of ``$\ldots$'' in subsequent lines are for typesetting only and indicate that the lines in
actual input should be joined into a single line.
The symbols \lcode{<d_} and \lcode{_d>} indicate the first derivatives of the bra and ket basis functions, 
respectively, as in Eq.~(\ref{eq:symmetricDerivative})  and \lcode{Q} is the 
placeholder for the coordinates $\varphi$, $\eta=\cos\th$, and $r$ at the respective positions in the tensor product. In practice,
for standard operators such as the Laplacian or partial derivatives $\ddx,\ddy$ and $\ddz$ short hand notation such as \lcode{<<Laplacian>>},
\lcode{<<D/DX>>} etc.\ can be used instead of the full definition.

\change{}{
Apart from possible right (\lcode{_d}) and left (\lcode{d_}) derivatives the string within the \lcode{<...>} is an 
algebraic expressions where \lcode{Q} is a placeholder for the coordinate in the respective tensor product. 
For the construction of admissible algebraic expressions see Sec.~\ref{sec:algebra}.
}

The code automatically infers from the input the Dirichlet boundary condition $\chi(r\!=\!0)\!=\!0$ and 
implements it by omitting the Lagrange polynomial $b^{0}_{0}(0)=1$ from the basis. For absorption, one
adds a special ``infinite'' element $[r\upp{N-1},\infty)$ with basis functions $b\upp{N-1}_{k}(r)$ 
based on the Gauss-Radau quadrature for Laguerre-type polynomials. This leaves the general structure of Eq.~(\ref{eq:expansion}) 
unchanged and provides for highly accurate and numerically efficient absorption, see discussion 
of irECS in Sec.~\ref{sec:tsurff}.

As an example we consider the Hydrogen atom, $V(|\vr|)=-\frac{1}{r}$, and the computation of 
photoelectron spectra
for a laser pulse with peak intensity of $2\times10^{14}W/cm^2$ at  central wave length of 
800 nm and a pulse duration of $5$ optical cycles at FWHM. (One optical cycle at circular frequency $\om$ is $2\pi/\om$.) 
In order do ensure the absence of 
any unphysical dc-component from the laser pulse, pulses are defined in terms of $\vA$ rather than $\vEf$ through 
pulse shape and polarization direction $\val(t)$ and the peak intensity $I_0$
\begin{equation}\label{eq:laserA}
\vA(t) = \val(t) \sqrt{\frac{I_0}{2\om^2}} \sin(\om t - \phi).
\end{equation}
The tRecX input for the pulse above is
\begin{lstlisting}
Laser: shape, I(W/cm2), FWHM, lambda(nm), phiCEO
      cos8,    2.e14, 5 OptCyc, 800.,      0
\end{lstlisting}
The \lcode{Laser:shape} and \lcode{FWHM} parameters determine $\val(t)$, which by default points into $z$-direction. Any desired 
polarization angle can be input with additional parameters. 
Shape \lcode{cos8} indicates a pulse envelope function $\cos^8$, which approximates a Gaussian pulse
but maintains strictly finite pulse duration, in this case about 3000 \au. 
At the carrier envelope offset phase $\phi=0$ the vector potential $|\vA|$ has a node at $t=0$. The field $\vEf(t)$ then has its peak
approximately at $t=0$ except for very short pulses, where the factorization into carrier and envelope becomes ill-defined and 
extra contributions from the time-derivative of $\val(t)$, see (\ref{eq:potA}), become non-negligible.

The discretization is specified in the form
\begin{lstlisting}
Axis:name,nCoefficients,...
     ...lower end,upper end,functions,order
 Phi,1
 Eta,30,-1,1, assocLegendre{Phi}
 Rn,80,  0,   40,polynomial,20
 Rn,20, 40,Infty,polExp[0.5]
\end{lstlisting}
This means that we use 30 angular momenta ($L_{\text{max}}=29$) and FE-DVR functions $b\upp{n}_{k}(r), n=0,1,2,3$ on  equal size sub-intervals of
$[0,40]$, each of order 20 with a total of 80=20$\times$4 coefficients. 
The FE-DVR basis $b\upp{4}_{k}$ starting at $40$, consists 20 polynomials with exponential damping $\exp(-0.5 r)$.  The single function
on the $\varphi$-coordinate is trivially constant and the associated Legendre functions here effectively 
reduce to the ordinary Legendre polynomials. Specifying the radial coordinate as \lcode{Rn} instructs the code to use
the Dirichlet boundary conditions at $0$ and a warning will be issued, if the basis does not start from $r=0$.
The remaining inputs for time-propagation and complex scaling, will be discussed in later examples.

\begin{figure}
\includegraphics[width=0.47\textwidth]{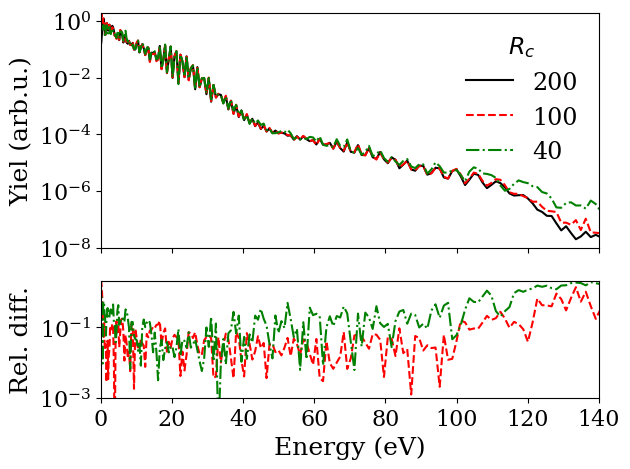}
\caption{\label{fig:fig1}
Dependence of calculated photo-emission spectra on the radius $R_c$ (c.f. Sec.~\ref{sec:tsurff}). 
Calculation for the Hydrogen atom with a laser pulse duration of 3 optical  cycles at wave length 800 nm and peak intensity 
$3\times10^{14}W/cm^2$. The relative differences of $\lesssim 10\%$ between the calculations arise, as the Coulomb tail
is effectively cut off at $R_c$. Numerical convergence is well below these differences. A script for producing this and similar 
plots directly from multiple tRecX outputs is provided with the code, see Sec.~\ref{sec:scripts}.
}
\end{figure}

At the given laser parameters tSurff was first demonstrated for a realistic scale problem 
in a prototype implementation \cite{Tao2012}. With tRecX results are obtained
within $\lesssim 3$ minutes on a modern CPU with the input listed above which delivers relative accuracies of the photo-electron spectra of about 
10$\sim$20\% in the main part of the spectrum, see Fig.~\ref{fig:fig1} and also discussion in \cite{Tao2012}. Computation times can be further reduced by 
parallelization, but gains of a factor $\lesssim 4$ on up to 8 cores remain moderate due to the small overall size of the problem,
see  Sec.~\ref{sec:scaling}.
A complete functional input with comments on the specific choices and on convergence 
is can be found in \lcode{tutorial/11}.

\subsection{The Helium atom in a strong laser field}
\label{sec:exampleHelium6d}
This much larger problem is used to illustrate the input of higher-dimensional and more complex discretizations that
contain basis constraints in the form of inter-dependencies between the coordinates. Also, with electron repulsion 
an operator appears that does not have tensor-product form. The \lcode{tutorial/23Helium3DSpectrum} elaborates further 
on the following by computing double-emission spectra, although for less demanding parameters.

The Hamiltonian of the Helium atom is
\begin{equation}\label{eq:hamHelium}
H_2(t) = H(t) \otimes \one + \one \otimes H(t) + \frac{1}{|\vr_1-\vr_2|},
\end{equation}
where $H(t)$ is the single-electron Hamiltonian from Eq.~(\ref{eq:tdse1}) with $V(r)=-\frac{2}{r}$. 
We generalize the expansion (\ref{eq:expansion}) to two electrons with the ansatz
\begin{equation}\label{eq:expansion2}
\Psi(\vr_1,\vr_2)=\sum_{m_1=-M}^{M}\sum_{m_2=-M}^{M} \sum_{l_1=|m_1|}^L \sum_{l_2=|m_2|}^L 
Y\upp{m_1}_{l_1}(\varphi_1,\th_1)Y\upp{m_2}_{l_2}(\varphi_2,\th_2)\chi^{m_1m_2}_{l_1l_2}(r_1,r_2)
\end{equation}
and the radial functions 
\begin{equation}
\chi^{m_1m_2}_{l_1l_2}(r_1,r_2)=
\sum_{n_1,n_2=0}^{N-1} \sum_{k_1,k_2=0}^{K-1} b\upp{n_1}_{k_1}(r_1) b\upp{n_2}_{k_2}(r_2)C^{m_1m_2n_1n_2}_{l_1l_2k_1k_2}(t).
\end{equation}
In a complete expansion for $\Psi(\vr_1,\vr_2)$ analogous to Eq.~(\ref{eq:expansion}), the 8-index coefficients appear within a hierarchy of 8 sums, with the number of indices related to the dimension of the problem. In tRecX, such a discretization 
can be specified by the lines
\begin{lstlisting}
#define BOX 40
#define ANG 20
#define NABS 15
Axis:name,nCoefficients,lower end,upper end,functions,order
 Phi1,  3,    0.,2*pi,expIm
 Eta1,  ANG, -1,1, assocLegendre{Phi1}
 Phi2,  3,0.,2*pi,expIm
 Eta2,  ANG,-1,1, assocLegendre{Phi2}
 Rn1,   20, 0.,10.,polynomial,20
 Rn1,   40, 10,BOX,polynomial,20
 Rn1,   NABS, BOX,Infty,polExp[1.]
 Rn2,   20, 0.,10.,polynomial,20
 Rn2,   40, 10,BOX,polynomial,20
 Rn2,   NABS, BOX,Infty,polExp[1.]
\end{lstlisting}
For convenience, the input files allow local macros, here used to define \lcode{ANG} as 20 for the number of angular momenta,
\lcode{BOX} for the simulation box size, and \lcode{NABS} for number of functions for absorption. The radial axes 
\lcode{Rn1} and \lcode{Rn2} are here cut into
three different regions, the section $[0,10]$ with 20 points, the region with lower density of 40 points on $[10,40]$, and the absorption region beyond 40. This choice accounts for the fact that higher momenta occur mostly near the nucleus, but, of course,
this intuition needs to be  verified by convergence studies. On the $\varphi_1$ and $\varphi_2$ coordinates we have the first three functions from the 
basis \lcode{expIm} which is defined as $\{1,e^{-i\varphi},e^{i\varphi},e^{-2i\varphi},e^{2i\varphi},\ldots\}$.

\subsubsection{Basis constraints and index hierarchy}
\label{sec:basisConstraint}
\begin{figure}
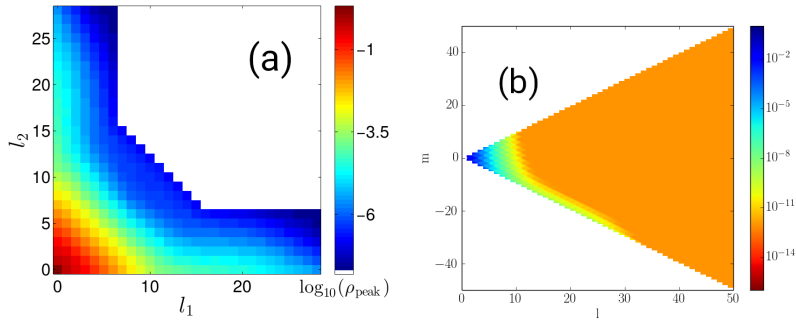

\includegraphics[width=0.44\textwidth]{constraint_lshape}    
\includegraphics[width=0.48\textwidth]{adaptivegrid}    
\caption{\label{fig:constraints}
Preponderance rules and the use of constraints. Panel (a) shows the maximal amplitudes of angular momentum pairs $(l_1,l_2)$ 
during laser ionization of Helium atom by a linearly polarized field. An \lcode{Lshape} constraint discards unneeded pairs. Reproduced from 
Ref.~\cite{zielinski16:doubleionization}. Panel (b): maximal amplitudes in the spherical waves $Y\upp{m}_l$ during ionization of a single-electron atom 
by a near-circularly polarized field. A pronounced preponderance for small $l+m$ is seen. Reproduced from Ref.~\cite{majety17:heElliptic}.
}
\end{figure}

Nominally, the above basis has a daunting size, given by the product of the size of each of the axes, which 
would be impractical for calculations. tRecX allows to impose constraints on the bases by letting the basis 
one hierarchy level depend on the preceding levels.
In fact, a first such constraint has tacitly been introduced by 
using the spherical harmonics $l_i\geq |m_i|$, where the $\aleg{m_i}{l_i}(\cos\th_i)$ depend on the value of $m_i$. 
For the given problem further constraints were added by the input
\begin{lstlisting}
BasisConstraint: axes,kind
Phi1.Phi2,M=0
Eta1.Eta2,Lshape[3;24]
\end{lstlisting}
The first line simply constrains the $z$-component of total angular momentum to $0=m_1+m_2$, which reduces the 6-dimensional 
problem to 5 dimensions, and, in our example reduces the basis size by a factor 3. The second constraint accounts for the fact 
that because of the particular dynamics of photo-ionization 
pairs of angular momenta $(l_1,l_2)$ with both values large do not occur and the basis can be constrained to 
an L-shaped region near the axes in the $l_1l_2$-plane, Fig.~\ref{fig:constraints}.  
Examples and numerical demonstration of such constraints can be found in \cite{zielinski16:doubleionization,majety17:heElliptic}.
This reduces the effective dimension to near 4. The possibility to flexibly 
impose constraints of this kind is one of the important features of the tree-structures in tRecX and has been used
extensively in applications.

As a result of the constraints the expansion coefficients $C$ no longer are the components of a tensor. 
Rather, the indices become inter-dependent, 
where we use the convention that any index can only depend on the indices to the left of it. While operator matrices cease to be
tensor products of matrices, the hierarchy of indices still allows efficient operator application, see Sec.~\ref{sec:quadratures}.

\change{}{
Presently only the \lcode{BasisConstraint}'s shown in the input documentation are available. Extension is easy for simple cases. 
That includes basic cases of spin, where spin can be added as a two-component \lcode{Vec} axis. A class must be derived
from \lcode{IndexConstraint} to handle the case. For implementation of non-local symmetries, such as multi-particle angular momentum 
or exchange symmetry, the use of constraints can become very complicated and direct implementation through explicitly symmetrized 
bases (to derive from \lcode{BasisAbstract}) may be more efficient both, in programming and computation. Note, however, that 
time-propagation dominantly depends on the sparsity and tensor-product structure of the operator matrix 
and only to a lesser degree on the length of the coefficient vectors. Also, non-locality of a symmetrized basis may deteriorate parallelization. These various aspects need to be considered when deciding for explicit implementation of symmetries. 
At present, tRecX mostly uses unsymmetrized, but in return sparse and factorizing representations.
}

\subsubsection{Coulomb repulsion}
\label{sec:coulombRepulsion}

Coulomb repulsion cannot be written as a finite tensor product and requires special treatment. We use a multipole
expansion and apply the radial part by multiplication on a quadrature grid. Although this can
be made exact within the given polynomial basis, it turns out that the approximate DVR quadrature 
does not compromise computation accuracy. Details of the scheme are given in Ref.~\cite{zielinski16:doubleionization} for 
finite elements, which can be readily transferred to FE-DVR now used by default in tRecX.

While tensor product operators can be defined through simple scripting, 
Coulomb repulsion is custom-implemented.
The Hamiltonian (\ref{eq:hamHelium}) can be specified as 
\begin{lstlisting}
Operator: hamiltonian=1/2<<Laplacian>>...
          ...-2.<<Coulomb>>+[[eeInt6DHelium]]
Operator: interaction=iLaserAz[t]<<D/DZ>>,
\end{lstlisting}
where the \lcode{<<...>>} are automatically converted to strings of the operator scripting discussed above, 
but \lcode{[[eeInt6DHelium]]} directs the code to a specialized operator class for electron repulsion. 
The separation into \lcode{hamiltonian} and \lcode{interaction} is for convenience only, internally the two
strings are merged into a single operator. Also note that the axes need not be 
given in exactly the sequence as shown in the example, if only one ensures that pieces belonging the same axis are in 
consecutive lines and that the functions on a given coordinate axis can only depend on coordinates specified above it: 
for example, \lcode{Phi2} must appear above the \lcode{Eta2} which carries the associated Legendre functions  \lcode{assocLegendre{Phi2}}. The sequence determines the layout of the indices of the $C$'s, where storage is such that lowest axis corresponds to 
the rightmost index, which runs fastest. Storage arrangement can be modified when defining the parallel layout, See.~\ref{sec:parallel}.

\subsection{Floquet calculation}
\label{sec:floquet}
The Floquet method converts a time-periodic problem into a stationary problem by discrete Fourier expansion in time. 
The resulting operator has continuous spectrum on the whole real axis, but underlying resonances can be accessed by complex scaling.
The \lcode{tutorial/90Floquet} shows how the method can be used within tRecX.

The TDSE for a single-electron system in a cw field polarized in $z$-direction is, in velocity gauge,
\begin{equation}
i\ddt \Psi(\vr,t)=\left[H_0 + i \sin(\om t) A_z\ddz\right]\Psi(\vr,t).
\end{equation}
The $\Psi(\vr,t)$ can be expanded into 
\begin{equation}
\Psi^\al(\vr,t)=e^{-i\ep^\al t}\Phi^\al(\vr,t),
\end{equation}
where the $\Phi^\al(\vr,t)=\Phi^\al(\vr,t+T)$ are strictly time-periodic and
in turn can be expanded into a discrete Fourier series
\begin{equation}\label{eq:floquetExpansion}
\Phi^\al(\vr,t)=\sum_{n=-\infty}^\infty  e^{in\om t}\Phi\upp{\al}_n(\vr),\quad \om=\frac{2\pi}{T}.
\end{equation}
Inserting into the TDSE and arranging the $\Phi^{\al}_n$ into a vector $\vPhi^\al$ one 
finds the eigenvalue equation
\begin{equation}
\mH_F \vPhi^\al = \ep^\al \vPhi^\al
\end{equation}
with
\begin{equation}\label{eq:floquetHam}
(\mH_F)_{mn} = \de_{nm} (H_0+ n\om) + \frac12(\de_{n,m+1}-\de_{n,m-1}) A_z\ddz 
\end{equation}
The Floquet Hamiltonian $H_F$ has the complete real axis as its continuous spectrum, into which
the bound states of $H_0$ are embedded. For non-zero $A_z$ all bound states experience an ac-Stark shift to
a resonance energy $E_r$ with a decay width $\Ga_r$. Upon complex scaling these two quantities appear
as complex eigenvalue $E_r-i\Ga/2$ of the complex scaled $H_F$.

We define a discretization for the expansion (\ref{eq:floquetExpansion}) as 
\begin{lstlisting}
Axis: name,nCoefficients,lower end,upper end,functions,order
 Vec,18
 Phi, 1
 Eta, 7,-1,1, assocLegendre{Phi}
 Rn, 60, 0.,BOX,polynomial,30
 Rn, 30, BOX,Infty,polExp[0.5]
\end{lstlisting}
where the first axis \lcode{Vec} labels a total of 18 Floquet blocks, i.e. the Fourier components $\Phi_n, n<18$. 
The Floquet Hamiltonian (\ref{eq:floquetHam}) is input as 
\begin{lstlisting}
#define KIN 1/2<<Laplacian>>-<1><1><1/Q+exp(-2.135*Q)/Q>
#define OM 0.1155<1><1><1>
#define INT A[I]/2(<delta[1]>-<delta[-1]>)<<D/DZ>>
Operator:hamiltonian=<Id>H0+<diagonal[Q-14]>OM+INT
\end{lstlisting}
where the \lcode{define} macros are used for better readability. The factor \lcode{<diagonal[Q-14]>} indicates 
a diagonal matrix for the first axis \lcode{Vec} with entries $(i-14)\de_{ij}$. 

\begin{figure}
\includegraphics[width=0.7\textwidth]{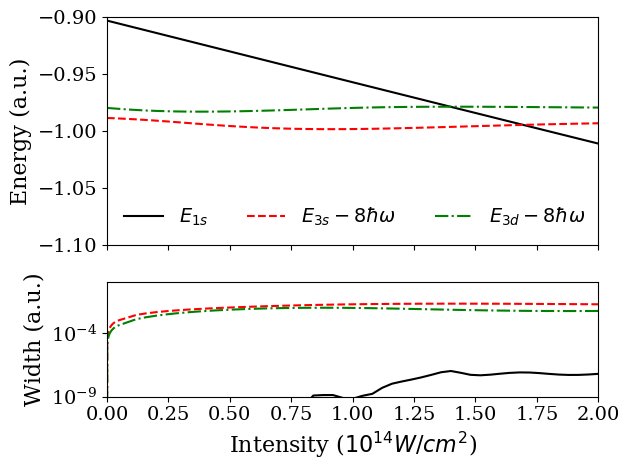}
\caption{\label{fig:floquet}
Floquet energies and widths as functions of laser intensity at photon energy 0.1115~\au ($\approx 400 nm$).
States can be identified by their initial field-free energies.
The crossings near $1.5\times10^{14}W/cm^2$ occur at  8-photon transitions from the ground to the 3s and 3d-states, respectively. Energies 
are \wrt the continuum threshold.}
\end{figure}

The potential $-(1+e^{-2.135 r})/r$ models the screened potential seen by one electron in a Helium atom. That model gives
qualitatively meaningful results for single-ionization processes and approximately reproduces the first few ground 
and excited state energies of the Helium atom. We use it to illustrate non-perturbative ac-Stark shifts and the resulting 
intensity-dependent $n$-photon Freeman resonances \cite{Freeman1987}.  
We  trace the resonance positions  $E_r-i\Ga_2$ as a functions of $A_z$ from field intensity $I=0$ 
to $2\times10^{14}W/cm^2$. The function \lcode{A[I]} that is used in the Hamiltonian string 
together with tracing range and step size are defined in the input as
\begin{lstlisting}
Trace: eigenvalues,from, to,   steps, function
        -0.903,    0, 2e14 W/cm2, 71, A[I]=sqrt(I)/0.1155
\end{lstlisting}
where $-0.903$ is the initial guess eigenvalue and the function $A[I]$ defines the conversion from
intensity to $A_z$ in \au for the given photon energy of $0.1155\,au \sim 3\,eV$. The eigenproblem is solved by 
inverse iteration and roots are selected for largest overlap with the preceding solution. Fig.~\ref{fig:floquet}
shows traces for ground and excited states, where crossings near intensities $1.5\times 10^{14}$ indicate an
8-photon resonance. These lead to characteristic structural changes in differential double emission spectra, as
discussed in Ref.~\cite{zhu20:doubleionization}.

\subsection{Model in two spatial dimensions}
\label{sec:exampleHe2d}

A popular model for inspirational studies in strong field physics is the ``two-dimensional Helium atom'' 
defined by the Hamiltonian 
\begin{equation}\label{eq:He2d}
H(x_1,x_2)= \sum_{i=1,2}\left[-\frac12\frac{\pde^2}{\pde x_i^2}+iA(t)\frac{\pde}{\pde x_i}-\frac{2}{\sqrt{x_i^2+a}}\right]+\frac{1}{\sqrt{(x_1-x_2)^2+b}},
\end{equation}
which with values $a=0.5$ and $b=0.3$ has a ground state energy of -2.88~\au and, remarkably, the exact single ionization threshold of -2~\au
The model owes its popularity to the fact that Fast Fourier Transform can be used for an efficient representation of the 
derivatives and  comparatively large spatial domains can be used to extract spectra by standard procedures.
In tRecX we use the model mostly for exploring numerical procedures and for testing new code, such as the first demonstration
of double-emission spectra in Ref.~\cite{scrinzi12:tsurff}. For a complete input example, see \lcode{tutorial/20Helium2d}

A Cartesian grid extending symmetrically around the origin is input as
\begin{lstlisting}
#define BOX 20
Axis:name,nCoefficients,lower end,upper end,functions,order
X1, 10,-Infty,-BOX.,polExp[0.5]
X1, 40,-BOX,BOX,polynomial,20
X1, 10, BOX,Infty,polExp[0.5]
X2, 10,-Infty,-BOX,polExp[0.5]
X2, 40,-BOX,BOX,polynomial,20
X2, 10, BOX,Infty,polExp[0.5]
\end{lstlisting}
The coordinates \lcode{X1,X2} illustrate the general tRecX feature that coordinates can be numbered. 
Equivalently one can use, e.g., the axis names
\lcode{X,Y}.
Complex scaling is input as 
\begin{lstlisting}
Absorption: kind, axis, theta, upper
ECS,X1,0.3,BOX
ECS,X2,0.3,BOX
\end{lstlisting}
with a complex scaling radius of $R_0=20$ at positive coordinates. The complex scaling radius at negative coordinates
defaults to $-R_0$, but can also be set explicitly by specifying a value for \lcode{Absorption:lower}.

Using input macros for brevity, the Hamiltonian is
\begin{lstlisting}
#define H1 (0.5<d_1_d>-<2/sqrt(Q*Q+0.5)>)
#define H2 H1<1>+<1>H1
Operator:hamiltonian=H2+<{}><1/sqrt(pow[2](X1-X2)+0.3)>
\end{lstlisting}
This illustrates how to define electron repulsion, which is a multiplicative operator that is 
not a tensor product \wrt $x_1$ and $x_2$: one defers the definition of the potential by putting a placeholder factor
\lcode{<{}>} until one reaches the hierarchy level of the
lowest coordinate axis, here \lcode{X2}. On that last level one defines the function using the axis names as the variables. 
Simple multi-dimensional
potentials can be input easily in this way. For more complicated dependencies one may consider writing a specialized class instead.
A larger class of general three-dimensional potentials is covered by the \lcode{Pot3d} discussed in section~\ref{sec:offCenter} below.

\begin{figure}
\includegraphics[width=0.95\textwidth]{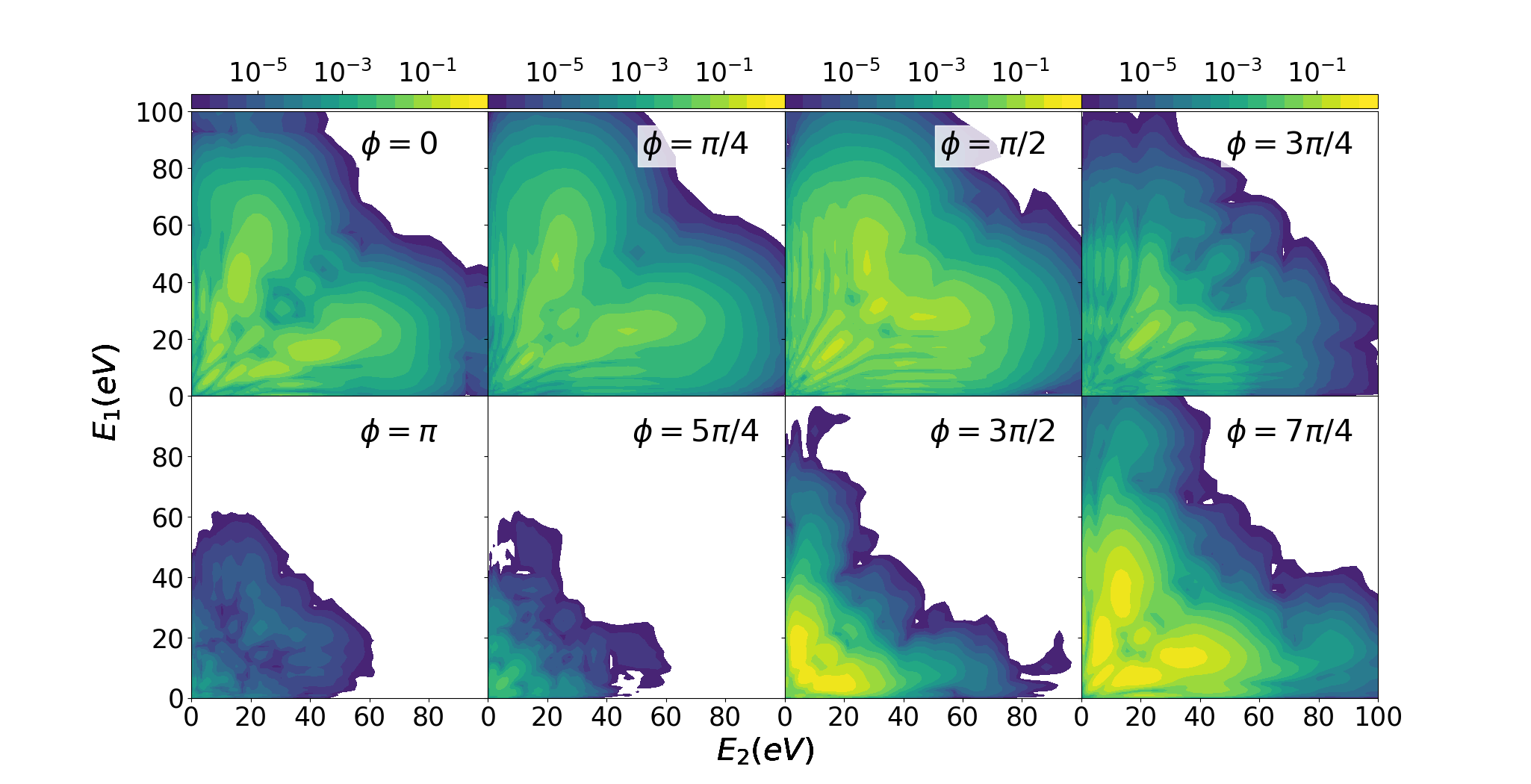}
\caption{\label{fig:spec_he2d}
Photo-electron spectra for the 2$\times$1-dimensional Helium model (\ref{eq:He2d}), dependence on the carrier-envelope phase $\phi$,
see Eq.~(\ref{eq:laserA}). All graphs share the same color code with normalization to overall maximum 1.
Only the quadrant where both electrons are emitted into the positive axis direction is shown. The pulse is single-cycle at wave length 800 nm and
intensity $3\times10^{14}W/cm^2$. Spectra strongly depend on $\phi$ because of the pronounced asymmetry of the single-cycle pulse. }
\end{figure}

Spectra for emission into the first quadrant $\RR^+\times \RR^+$ can be computed by inputs analogous to the full 6-dimensional case.
Other quadrants are not supported at present, but spectra can be obtained by computations with 
reflected coordinate axes $(x_1,x_2)\to(\pm x_1,\pm x_2)$. Fig.~\ref{fig:spec_he2d} shows the dependence of spectra on the carrier-envelope phase
$\phi$, Eq.~(\ref{eq:laserA}), for a single-cycle pulse.

\subsection{Molecular model}
\label{sec:exampleMolecule}
In tRecX one can use hybrid bases where different types of basis functions are combined to discretize the same space.
A typical example is the haCC method \cite{majety15:hacc} for molecules in strong fields, 
which combines a Gaussian-based CI with the numerical basis described 
above. Another example is a multi-center basis, where spherical bases with different centers are combined. 

For the introduction of the concept of hybrid bases we use a model that is popular in strong field physics, realized in 
\lcode{tutorial/221CO2Free}.
In that type of model one assumes that a single or a few bound states $\{|\al\r,\al=\range{A}\}$ of some complicated Hamiltonian $H_a$ 
are essential, but the strong field dynamics on the rest of the space can be described by a simplified Hamiltonian $H_b$ with the total Hamiltonian
\begin{equation}\label{eq:hamSubspace}
H(t)=P H_a P+QH_bQ + i\vA(t)\cdot\vna,\quad\text{and}\quad Q=(1-P), P=\sum_{\al=0}^{A-1} |\al\r\l\al|.
\end{equation}
For $H_b$ one typically uses free motion or motion in a Coulomb field. 
Note that the interaction among the $|\al\r$ states and between $|\al\r$ and the rest of the space is taken fully 
into account in $H(t)$. With a single bound state $A=1$ and $H_b=-\frac12\De$ this is very nearly the so-called 
``strong field approximation'' \cite{Lewenstein1994}, which is behind much of the theoretical understanding of strong 
field physics. Hamiltonian (\ref{eq:hamSubspace}) was used to investigate attosecond ($1\,as=10^{-18}s$) delays in photo-emission from $CO_2$. 
We choose a highly simplified $CO_2$ model Hamiltonian
\begin{equation}\label{eq:CO2model}
H_a=-\frac12\De 
- \frac{\ga(1+5e^{-r/c})}{|\vr|} 
- \frac{(1-\ga)(1+7e^{-|\vr+\vb|/a})}{2|\vr+\vb|}
- \frac{(1-\ga)(1+7e^{-|\vr-\vb|/a})}{2|\vr-\vb|},
\end{equation}
where $\ga=0.5$ parameterizes the distribution of charge between the $C$ and $O$ atoms and screening was chosen as $c=3,a=1.73$.
The O-atoms are located along the $z$-axis at the equilibrium $C\!-\!O$ bond length $\vb=(0,0,2.197\,au)$.
With that one finds a $\Pi$-gerade state at the $CO_2$ HOMO energy of $\approx-0.51 \au$. For the purpose of this
study it suffices to compute the eigenstates $|\al\r$ of $H_a$ in a single-center expansion. We pick the HOMO and 
the next higher $\Sigma$-state as follows:
\begin{lstlisting}
#define CO2 <1><{}><CO2Pot[BOX,GAM,CSCR,ASCR](Eta,Rn)>
#define HAM (1/2<<Laplacian>>+CO2)
Axis: subset,name,nCoefficients,lower end, upper end,...
    ...functions,order
  Subspace,Orbital,2,7,,Eigenbasis[HAM:Complement]
Complement,    Phi,  7
          ,    Eta, 10,-1,    1,assocLegendre{Phi}
          ,     Rn, 80, 0,   40,polynomial,20
          ,     Rn, 20,40,Infty,polExp[0.5]    
\end{lstlisting}
The potential parameters \lcode{BOX,GAM,CSCR,ASCR} are set by \lcode{define}'s. The function
itself was hard-coded into tRecX for efficiency, although it can be, in principle, written as in the example
of Sec.~\ref{sec:exampleHe2d}.
The additional input \lcode{subset} separately specifies the discretization on a \lcode{Subspace} and its
\lcode{Complement}. The basis on the subspace are two \lcode{Orbital}s $\{\Phi_0(\vr),\Phi_1(\vr)\}$, which 
are three-dimensional \lcode{Eigenbasis}  functions of the Hamiltonian \lcode{HAM}, which are 
computed in the discretization defined in the \lcode{subset} named \lcode{Complement}. 
\lcode{Complement} is a standard spherical expansion. The axial symmetry around $z$ is broken
by the field, which is why a total of 7 $\varphi$-functions $m=1,\pm1,\pm2$ are used. This suffices as we only study two-photon
transitions in the perturbative limit. 

The dipole field of the laser is specified by the fundamental $\om$ and its 13th and 15th harmonic as
\begin{equation}
\vA(t)=\uep A_f \cos^2(\frac{t}{\tau_0})\sin(\om t +\phi)+ \uep A_h \cos^4(\frac{t}{\tau_H})[\sin(13\om t)+\sin(15\om t)]
\end{equation}
with the polarization vector $\uep$ in the $xz$-plane.
The field is input in terms of peak intensities and FWHM as 
\begin{lstlisting}
Laser:shape,I(W/cm2),FWHM, lambda(nm),phiCEO, polarAngle
       cos2,  1e10, 4 OptCyc, 800,      pi/2,  45
       cos4,  1e11, 3 OptCyc, 800/13,   0,     45
       cos4,  1e11, 3 OptCyc, 800/15,   0,     45
\end{lstlisting}
Here  \lcode{pi/2}  for \lcode{phiCEO} at the fundamental means that node of the 
fundamental field falls onto the peak intensity of the harmonics. Note that \lcode{OptCyc} is
\wrt to the first wave length in the list. A warning issued by the code will remind the user of this fact.

The Hamiltonian (\ref{eq:hamSubspace}) is specified as
\begin{lstlisting}
Operator: hamiltonian=<0,0>HAM+<1,1>(1/2<<Laplacian>>)
Operator: interaction=<allOnes>...
             ...(iLaserAx[t]<<D/DX>>+iLaserAz[t]<<D/DZ>>)
\end{lstlisting}
The factor \lcode{<allOnes>} is a matrix filled with $1$'s. It refers to the hybrid ``coordinate'' axis
\lcode{Subspace&Complement} and indicates that all sub-blocks of the interaction on the subspace and its complement are to be computed:
\begin{equation}
H_I=i\vA\cdot\vna=PH_IP+PH_I(1-P)+(1-P)H_IP+(1-P)H_I(1-P).
\end{equation}
Also note that polarization is no longer along the $z$-axis but rather in the $xz$-plane, which is why $x$ and $z$-components of 
the dipole interaction are both present.

\change{}{Further possibilities to set up the Hamiltonian are to select more and different orbitals in the \lcode{subset} space. 
Also, numerical values of small matrices for the construction of Hamiltonian and interaction can be specified in the input, see
\lcode{tutorial/221} for an illustration.}

\subsubsection{Orientation dependence of time-delays in photo-electron emission}
Delays in the laser-emission of electrons have drawn 
some attention as possible indicators of a delay in tunneling emission (see, e.g., \cite{pfeiffer12:attoclock,torlina13:ionization-time}), which
may occur on the time scale of attoseconds. In order to correctly pose the question, one must disentangle
any possible such delay from delays not related to tunneling that are well known to appear in scattering after emission. 
The model above allows to give meaning to the notion of ``scattering after emission'' by restricting the action of the binding potential
to the initial state and use the free particle Hamiltonian everywhere outside the bound initial state. This can be compared to the full problem,
or a partially restricted problem, e.g. using only the short range part of the molecular potential or motion in the Coulomb field instead of 
than free motion. The experimental definition of delay is related to a beat in a so-called RABITT spectrogram, 
see, e.g. \cite{scrinzi06:review} for a general discussion of attosecond techniques. 

\begin{figure}
\includegraphics[width=0.5\textwidth]{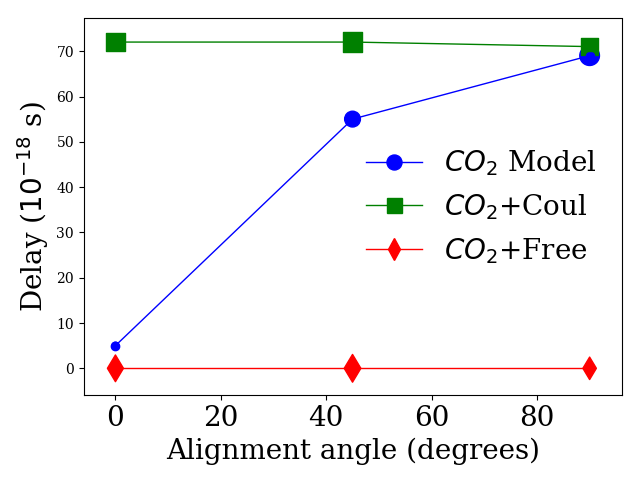}
\caption{\label{fig:CO2_alignment} 
Dependence of RABITT emission delays on alignment between of the molecular axis and laser polarization direction.
The three calculations are for the $CO_2$ single-electron model (\ref{eq:CO2model}, dots) and the Hamiltonian
(\ref{eq:hamSubspace}) with $H_b$ the free motion (diamonds) and motion in the Coulomb field (squares), respectively.
Size of the dots indicates emission yield. Alignment-dependent delays are largely are due to scattering in 
the molecular potential. Delays by the Coulomb potential are nearly independent of alignment.
}
\end{figure}

Here we only illustrate the use of tRecX for comparing alternative models
within the same computational framework without any deeper discussion of the underlying physics. 
Fig.~\ref{fig:CO2_alignment} shows RABITT delays computed with three different models,
the full single-electron Hamiltonian (\ref{eq:CO2model}), the strong field-like approximation Eq.~(\ref{eq:hamSubspace}) with 
free motion $H_b=-\De/2$ outside the ground state, and Coulomb scattering $H_b= -\De/2-1/r$. If there were any dependence of the delays 
on the alignment of the laser field with the molecular axis, this would be considered as an effect of 
tunneling through the orientation-dependent barrier. While the full model shows strong orientation dependence, no such 
effect is seen with free motion or motion in the Coulomb field. The conclusion from this simple study is that any possible
effects of tunneling delays would be completely dominated by delays incurring after emission.

\subsection{Multi-center bases}
\label{sec:offCenter}
When a system has singularities at several points in space the use of a multi-center basis is advisable. The  
\lcode{tutorial/510OffCenterScatter} was used as the starting point for the calculations published in \cite{jelovina20}.

We consider one scatterer at some larger distance from the origin. Such a potential cannot be written as tensor product \wrt the original polar coordinates, but rather is treated as a general three-dimensional potential which is input in a category
\lcode{Pot3d}. It is referenced in the operator definition as the special operator \lcode{[[Pot3d]]}. 
An off-center radial potential can be specified by the Cartesian coordinates of its origin 
and an algebra string for the radial function, as for $-1/\sqrt{|\vr-\vr_0|}$ with $\vr_0=(0,0,65)$ in 
\begin{lstlisting}
Pot3d: potential=radial[0,0,65,-1/Q]
\end{lstlisting}
A matching off-center basis with the spherical harmonics $l\leq 2$ is specified as
\begin{lstlisting}
PolarOffCenter:radius=5,origin=[0,0,65],Lmax=2,Mmax=0,Nmax=5
\end{lstlisting}
which uses polynomials of degree 4 on a sphere of \lcode{radius=5} around $\vr_0$. 
With the center placed on the $z$-axis, we have axial symmetry around the $z$-axis and $m$-quantum numbers remain conserved. 
In that case one may constrain the off-center basis to $m=0$, as in the example above. 

That basis is to be combined with a standard spherical basis centered at the origin into a hybrid basis as in
\begin{lstlisting}
Axis: subset,name,functions,nCoefficients,...
                ...lower end,upper end,order
   Off,Ndim,PolarOffCenter
Center,Phi,,1
      ,Eta,assocLegendre{Phi},4, -1, 1
      , Rn, polynomial,     160,  0,80,10
      , Rn, polExp[1.],      20, 80,Infty
\end{lstlisting}
The off-center potential and the off-center basis both break rotational symmetry and cause partial fill-in of overlap and operator matrices. Note that here, different from Sec.~\ref{sec:exampleMolecule}, the bases of the two subsets are not orthogonal. The inverse of the overlap is applied through a specialized class that implements the Woodbury formula for low-dimensional updates of an inverse (cf. Sec.~\ref{sec:otherOperators}). Possible linear dependency and ill-conditioning of the overlap is monitored, but does not usually pose a problem for a rather well-localized off-center basis as in this example. 

The fill-in of operator matrices occurs where the off-center functions overlap with the radial sections of the origin-centered basis. 
For that reason it is recommended to minimize the number of radial elements $[r^n,r^{n+1}]$ where the \lcode{Center}-basis overlaps 
with the \lcode{Off}-basis. 
In the given example, the off-center basis has overlap with two radial sections $r\in[60,70]$. 

Operators must be defined with respect to the \lcode{Center} discretization, as in 
\begin{lstlisting}
Operator: hamiltonian=0.5<allOnes><<Laplacian>>...
             ...-<allOnes><<Coulomb>>+[[Pot3d]]
\end{lstlisting}
The factor \lcode{<allOnes>} translates into a 2$\times$2-matrix filled with 1's for the hybrid \lcode{Off&Center} axis. 
This indicates that matrix elements between the basis functions of the two subsets are non-zero, when the functions overlap spatially. Matrix elements between all parts of the basis are computed using quadratures. When any of the functions is off-center, three-dimensional quadrature for 
the off-center basis employed, as typically the center-basis is smooth across the support of the off-center basis, e.g. a small solid angle
from the sphere times a polynomial in $r$.

\subsection{Further tutorials}
\change{}{
Except for the representative examples above, all standard features of the code are demonstrated with 
inputs in the \lcode{tutorial} subdirectories. The inputs included there at the time of writing are presented with brief
descriptions in Table.~\ref{tab:tutorials}.
}
\begin{table}
\begin{tabular}{ll}
00HarmonicOsc1 &  1d-HO --- discretization and eigenvalues\\
01HarmonicOsc2 &  2d-HO --- combine two discretization axes\\
02HarmonicOscPolar & 3d-HO --- polar coordinate, input of operators \\
03HydrogenPolar & 3d-hydrogen atom --- plot densities \\
04Hyd1d &  ``1d-hydrogen atom'' --- model and numerics\\
05irECS &  irECS for the 1d hydrogen atom\\
06TimeProp &  Basics of time propagation \\
07HighHarmonicGeneration & \ref{sec:stiffness} High harmonic spectra \\
08Hyd1dSpectrum & Photoelectron spectrum (1d) \\
09HydrogenSpectrum & 3d hydrogen: photo-electron spectrum at 20 nm \\
10IRSpectrum &  3d hydrogen: photo-electron spectrum at 800 nm\\
11shortPulseIR &  \ref{sec:exampleSingle} 3d hydrogen: strong IR pulse\\
12IRlongPulse &   3d hydrogen: strong and long IR pulse \\
13Circular400nm &  \ref{sec:adaptive} Circular polarization, 400 nm wave length \\
14Circular400nmLong & Circular polarization, longer pulse \\
15Elliptic400nm &  Elliptic polarization\\
15TayloredField &  Two-color field at general polarization \\
16RotatingFrame &  Rotating frame: photoemission at 400 nm\\
a16Circular800nm & Rotating frame:  photoemission at 800 nm \\
17MixedGauge & Mixed gauge, field as tutorial 10 \\ 
19TwoColorHarmonics &  Calculation of harmonics, 2-color driver \\
20Helium2d & \ref{sec:exampleHe2d} Double-emission: 1+1-dimensional He \\
21Helium2dIR & \ref{sec:stiffness} 1+1-dimensional He, IR pulse\\
22Helium6d &  Ground state of the He atom\\
23Helium3DSpectrum & \ref{sec:exampleHelium6d} Double emission from He \\
51ParabolicHarmonic & Harmonic oscillator in parabolic coordinates \\
70RabittDelays &  Attosecond RABITT delay calculation \\
90Floquet &  \ref{sec:floquet} Floquet calculation.\\
110Pot2d &  Variants of inputting 2d potentials\\
111Pot2dCO2 & A simple 2d CO2 model \\
220HybridSubspace &  Hybrid of orbital and numerical basis \\
221CO2Free &  \ref{sec:exampleMolecule} Strong-field-approximation type model \\
510offCenterScatter & \ref{sec:offCenter} Combine spherical with off-center basis\\
\end{tabular}
\caption{\label{tab:tutorials}
\change{}{List of tutorials supplied with the code. Tutorials 00-10 provide a basic introduction of inputs. 
Tutorials referenced in the present paper have links to the respective sections.}}
\end{table}

\section{Methods and general framework}

\subsection{irECS and tSurff}
\label{sec:tsurff}
Strong field problems involve photo-emission all the way to total ionization of the initial system. Pulse durations are
long on the atomic time scale and the momentum spectrum can be very broad. In this situation efficient absorption of outgoing
flux is provided by ``infinite range exterior complex scaling'' (irECS) \cite{Scrinzi2010}. Complex scaling is an analytic
continuation technique for Schr\"odinger operators by which the continuous energy spectrum is rotated around the 
single or multiple continuum thresholds into the lower complex plane leading to damping of the continuous energies in forward 
time-evolution. Bound state energies remain unaffected by the transformation and a new class of discrete eigenvalues $W_r=E_r-i\Ga_r/2$
appears that belong to square-integrable resonance states at energies $E_r$ with decay widths $\Ga_r$. 

The transformation is achieved by scaling the coordinates $\vr\to e^{i\th} \vr$. If the scaling is only applied
outside a finite radius $R_0$ one speaks of exterior complex scaling (ECS). As a consequence of analyticity, 
exterior complex scaling leaves the solution in the region $r\leq R_0$ strictly unchanged and allows 
direct physics interpretation --- it is a perfect absorber. The usual discretization errors arise  but any
dependence on the complex scaling angle $\th$ can be reduced to machine precision and in that sense there are no adjustable
parameters. The choice of $\th$ does matter for efficiency with $\th\sim\pi/10-\pi/6$ usually giving best results.
A particularly efficient discretization is used in irECS with exponentially damped polynomials in the scaled region $r>R_0$
\begin{equation}
b_k(r)=L_k(r)e^{-\al r}.
\end{equation}
The $L_k$ is any set of orthogonal or sufficiently well-conditioned polynomials such as Laguerre or Lagrange polynomials. 
In tRecX, we use for $L_k$ the Lagrange polynomials at the Radau quadrature points for the weight $e^{-2\al r}$, which is
a DVR basis (Sec.~\ref{sec:basisDVR}). The rationale of this discretization is to simultaneously accommodate short and long wave lengths:
short wave-lengths require finer sampling but get damped by complex scaling over a short range. Long wave lengths penetrate
deeper into the absorbing region, but need fewer discretization functions over the range. This discretization reduces the number
of functions needed for absorption per coordinate by factors $\lesssim 4$ from the already efficient absorption by ECS, 
an advantage that plays out especially in higher dimensions.

The input of the irECS parameters for the example of the two radial coordinates in the Helium problem of Sec.~\ref{sec:exampleHelium6d} is
\begin{lstlisting}
Absorption: kind, axis, theta, upper
ECS,Rn1,0.3,20
ECS,Rn2,0.3,20
\end{lstlisting}
The name \lcode{upper} indicates the complex scaling radius $R_0$ for interval  $[R_0,\infty)$. 
For Cartesian coordinates (Sec.~\ref{sec:exampleHe2d}) one also needs absorption towards negative infinity $(-\infty,X_{-}]$ which defaults to 
$X_{-}=-X_{+}$, but can be set independently by \lcode{lower} if so desired.
The exponentially damped functions are chosen with the axes, as shown in the applications of Sec.~\ref{sec:applications}.

The time-dependent surface flux (tSurff) method constructs spectra from the flux through a surface at
some sufficiently large radius $R_c$. It is specific for the dipole approximation used in 
laser-ionization that momenta will get modified also after they pass any remote surface. This can be taken into account 
if one has an analytic solution for the time-evolution outside $R_c$. With $R_c$ large enough for neglecting the 
potentials, these are the Volkov solutions for electronic motion in a dipole field, here given in velocity gauge and 
$\de$-normalized \change{}{(\wrt $\vk$) }
\begin{equation}
\chi^V_\vk(\vr,t)=(2\pi)\inv{3/2} e^{-i\Phi(\vk,t)}e^{i\vk\vr},
\end{equation}
with the $\vk$-dependent Volkov phases $\Phi(\vk,t)=\int_0^t d\tau [\vk-\vA(\tau)]^2/2$.
With these the complete spectral amplitude at a given $\vk$ can be written as an integral over the surface and time
\begin{equation}\label{eq:spectralAmplitudes}
b(\vk,T)=\int_{T_0}^T  \l \chi^V_\vk(t)|[-\frac12(-i\vna-\vA(t))^2,h(r-R_c)]|\Psi(t)\r dt
\end{equation}
As $h$ is the Heaviside function, the commutator leads to $\de$-functions at $r=R_c$ and the integral is only over 
the surface. $T_0$ is the begin time of the pulse, and $T$ is some time large enough such that all relevant flux
has passed $R_c$. 
The scheme written here for the single-particle emission can be generalized to the emission of two or more particles.
In tRecX, the general form is implemented, but in practice three-particle emission has not been studied 
for reasons of problem size. Further details on the tSurff method can be found in Refs.~\cite{Tao2012,scrinzi12:tsurff}.

tRecX computes values and derivatives of $\Psi(t)$ on the surface and saves them to disk. In a second sweep, the integral 
(\ref{eq:spectralAmplitudes}) for the spectral 
amplitudes $b(\vk,T)$ is computed. For multi-particle spectra the process is recursively iterated. One can specify the 
desired grid for $\vk$ using the input category \lcode{Spectrum} with a choice of points and optionally a momentum range. If
\lcode{Spectrum} is found, tRecX automatically initiates the amplitude computation. Alternatively, one can restart tRecX  with the
output directory as input and the momentum grid specified by command line parameters.

tSurff and irECS are the two defining techniques of tRecX which also have phonetically inspired the name as
tRecX=tSurff+irECS. An alternative interpretation of the acronym is related to the recursive discretization discussed 
below in Sec.~\ref{sec:discretization}.

\subsubsection{Discretization of complex scaled operators}
\label{sec:complexSymmetric}
The matrix representing a complex scaled Hamiltonian is non-hermitian and has the desired complex eigenvalues.
If one uses strictly real basis functions, the matrix for the unscaled Hamiltonian will usually be real.
In that case, the Hamiltonian matrix will become {\em complex symmetric} upon scaling, i.e. $\mH_{ij}=\mH_{ji}$ 
{\em without} complex conjugation.
This is a computationally useful property: the right eigenvectors of $\mH$ are identical to the left-eigenvectors
\beq
\mH \vC_n  = \vC_n W_n \Leftrightarrow \vC^T_n \mH = W_n \vC^T_n
\eeq
and the eigenvectors $\vC_n$ can be selected to be pseudo-orthonormal
\beq
\vC_m^T\vC_n = \de_{mn}.
\eeq

A modification of that general approach is used for irECS: one starts from strictly real basis functions on the \rhs, 
but in the scaled region $r\geq R_0$ these are multiplied by a complex factor. This creates a complex-valued discontinuity 
of the logarithmic derivative in \rhs basis at $R_0$, that is required by ECS. The analogous,
but complex conjugated discontinuity is required for the \lhs basis. Mathematical and implementation details of this realization of ECS, 
and its numerical advantages compared to commonly used alternatives are discussed in Refs.~\cite{Scrinzi2010,weinmuller17:dvr}.
With the \lhs differing from the \rhs basis, also the overlap matrix becomes complex symmetric rather than hermitian. However, algorithms
remain unchanged from the hermitian case, if some care is taken to properly use transposed instead of adjoint matrices and vectors. For example, a
pseudo-Schmidt-orthonormalization can be performed if only one replaces the standard scalar product with its complex-symmetric counterpart
$\vC\adj\vC\to \vC^T\vC$, and even a pseudo-Cholesky decomposition exists and is used. 
In tRecX, a keyword \lcode{pseudo} indicates that the unconjugated rather than standard operation 
is performed.

\change{}{
The above approach keeps its simplicity only, when the original Hamiltonian matrix \wrt the chosen basis is real. 
As the resulting complex symmetry of the complex scaled problem simplifies and accelerates algorithms, 
an effort should be made to find such a representation. In fact, at present tRecX does not reliably support cases, 
where the original unscaled matrix would be non-hermitian. 
While non-real hermitian matrices cannot ruled out in general, in all applications shown here matrices are indeed real. 
For example, in the Floquet problem a real matrix is obtained by defining a factor $i^n$ into the basis of 
the $n$'th block, which results in the overall hermitian definition for the interaction as 
\lcode{(<delta[1]>-<delta[-1]>)<<D/DZ>>} in Sec.~$\ref{sec:floquet}$.
}

\subsection{Recursive discretization}
\label{sec:discretization}

The organization of operators, wave-functions, expansion coefficients, basis sets, and multi-indices
in trees is central to the design of tRecX. Trees are used to 
recursively generate the objects and in virtually all other algorithms. This makes
the code largely independent of specific coordinate systems and dimensions and allows to handle all 
multi-dimensional expansions of the examples above within the same scheme. Program uniformity is ensured by deriving
all trees from a template abstract base \lcode{class Tree}, Sec.~\ref{sec:classTree}.

\subsubsection{Wave function expansion}
We denote the $L$-tuple of all coordinates by $Q=Q^0=(\rangesup{q}{L})$ and the sub-tuple starting at $l$ by $Q^l=(\rangeSup{q}{l}{L-1})$.
There is some flexibility as to what is considered as a ``coordinate'': on the 
one hand, the finite-element index $n$ in Eq.~(\ref{eq:expansion}) can assume the role of a coordinate, but also
all three spatial coordinates $\vr$ of the orbitals $\Phi_\al(\vr)$ in the hybrid discretization of Sec.~\ref{sec:exampleMolecule}
can be subsumed in a single $q^l:=\vr$.

One or several sets of basis functions $\bas\upp{J_l}=\left(b\upp{J_l}_{0}(q^l),b\upp{J_l}_{1}(q^l),\ldots\right)$ are defined for a coordinate $q^l$,
where we arranged the set as a row vector, indicated by the underscore. The multi-index $J_l=(j_0,\ldots,j_{l-1})$ unites the labels $j_k$
of all functions $b\upp{J_{k}}_{j_k}(q^k), k<l$ preceding the basis set $\bas\upp{J_l}(q^l)$. In that way $\bas\upp{J_l}$ 
can be made to depend on the sequence of basis functions preceding it in the coordinate hierarchy.  The tuple $J=J_L$ is the complete set of  
indices for a single expansion coefficient $C^J=C^{\rangesub{j}{L}}$. The basis function matching $C^J$ is $B_{J}(Q)=\prod_{l=0}^{L-1} b^{J_l}_{j_l}(q^l)$. 
As with coordinates, we use the word ``basis function'' in a rather wide sense: basis functions in the proper sense
are trigonometric functions, associated Legendre functions, or the Lagrange polynomials for FE-DVR discretization, etc.\ 
but we also consider Kronecker $\delta$: $b_{j}(n)=\delta_{jn}$ as a ``basis function'' for a discrete
index $n$, e.g. the photon index in the Floquet model of Sec.~\ref{sec:floquet}. By that principle 
all discretization methods are treated uniformly in tRecX.

All basis expansions discussed in Sec.~\ref{sec:applications} fit into the scheme, e.g. Eqs.~(\ref{eq:expansion}) and (\ref{eq:expansion2}).
It is important to note that, while individual functions $B_{J}$ are products of functions of the coordinates, the total basis $\Bas$ (again 
considered as a row vector of $B_{J}$'s) is {\em not } a product basis because of the dependence of factor functions $b\upp{J_l}_k$ 
on the complete preceding hierarchy. A well known set of two-dimensional functions with this structure
are the spherical harmonics $Y\upp{m}_l(\phi,\th)\propto e^{im\phi}\aleg{m}{l}(\cos\th)$. Another example for the hierarchical dependence in the
products is the implementation of angular constraints as discussed in Sec.~\ref{sec:exampleHelium6d}. 

With the above definitions, the wave function for a given tuple of coordinates $Q^l$ is expanded recursively as 
\begin{equation}\label{eq:recursivePsi}
\Psi^{J_{l}}(Q^l)=\sum_{j=0}^{K_{J_l}-1} b^{J_l}_{j}(q^l) \Psi^{J_{l}j}(Q^{l+1}),
\end{equation} 
where we use the notation $J_lj:=(j_0,\ldots,j_{l-1},j)$.
$\Psi(Q):=\Psi^{J_0}(Q^0)$ is the complete wave function at the point $Q$ and $\Psi^{J_L}=:C^{J_L}=C^{\rangesub{j}{L}}$ is a vector of length one, i.e. the complex valued expansion coefficient at the multi-index $J_l=(\rangesub{j}{L})$. 
The recursion (\ref{eq:recursivePsi}) defines the discretization as a tree whose nodes are labeled by an index $J_l$, as in 
Fig.~\ref{fig:indexTree}. The subtree starting at $J_l$ defines a multi-coordinate wave function component $\Psi\upp{J_{l}}(Q^l)$.
Every node hosts a basis $\bas\upp{J_l}=\left(b^{J_l}_j(q^l),\, j=0,\ldots,K_{J_l}-1\right)$  and each function of the basis
connects to one branch of the node. 
Both, the number $K_{J_l}$ of basis functions $b^{J_l}_j$ and their kind can be different on every node, as, e.g., 
for the associated Legendre functions $b^{J_2}_{l-|m|}=\aleg{m}{l}$ in Fig.~\ref{fig:indexTree}.
Usually basis sets at given level $l$ have equal coordinate $q^l$. An exception are hybrid discretizations as in Secs.~\ref{sec:exampleMolecule} and \ref{sec:offCenter}. In Fig.~\ref{fig:indexTree}, on level $l=1$ the node at $J_1=(0)$ hosts three-dimensional eigenfunctions 
$\bas\upp0=\left(\Phi_0(\vr).\Phi_1(\vr)\right)$, while its neighbor at $J_1=(1)$ has the node-basis 
$\bas\upp1=\left(1,e^{-i\phi},e^{i\phi}\right)$.

\begin{figure}
\end{figure}
\begin{figure}
\includegraphics[width=\textwidth]{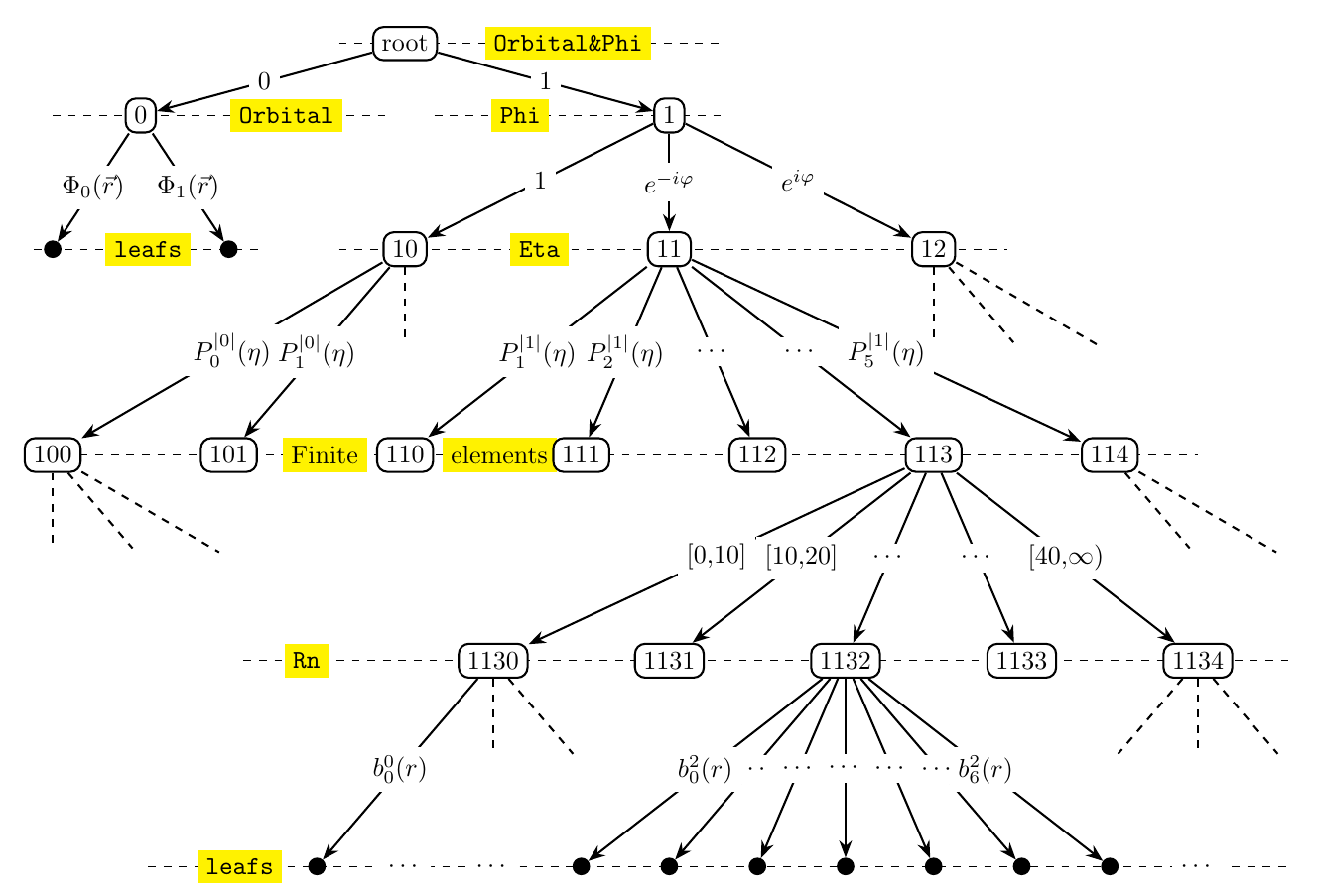}
\caption{\label{fig:indexTree}
Index tree for the hybrid discretization of Sec.~\ref{sec:exampleMolecule} (abbreviated). Axis names
are indicated in yellow. Nodes on the respective levels are labeled by $J_l$, factor basis functions connect a node to 
the next-lower level.}
\end{figure}

\subsubsection{Coefficients and operator matrices}

The recursive hierarchy is also reflected in the expansion coefficients. Every subtree wave function $\Psi^{J_l}$ 
is associated with a vector of expansion coefficients $\Coef^{J_l}$. The overline indicates a column vector and emphasizes 
its duality to the basis $\Bas\upp{J_l}$. $\Coef\upp{J_l}$ is the 
direct sum of the coefficient vectors at $l+1$: 
\begin{equation}\label{eq:recursiveC}
\Coef^{J_l}=\bma \Coef^{J_l0}\\\Coef^{J_l1}\\\vdots\\\Coef^{J_lK_{J_l}}\ema.
\end{equation}
The recursion can be phrased as
``a coefficient vector is a vector of coefficient vectors''. 

Finally, the recursive hierarchy of the overall multi-dimensional basis set belonging to $\Coef$ 
can be exploited for the computation of the operator matrices and in matrix-vector multiplication. 
The row-vector of multi-dimensional basis functions $\Bas\upp{J_l}(Q^l)$ for the subtree $J_l$ is defined recursively as
\begin{equation}\label{eq:recursiveBasis}
\Bas^{J_{l}}(Q^l)= \left( b^{J_l}_{0}\Bas\upp{J_{l}0},\,b^{J_l}_{1} \Bas\upp{J_{l}1},\ldots,\, b^{J_l}_{K_{J_l}-1} \Bas\upp{J_{l}K_{J_l}-1}\right).
\end{equation}
The recursion starts from $B^{J_L}\equiv1\forall J_L=(\rangesub{j}{L})$ and $\Bas^{J_0}=\Bas$ is the complete multi-dimensional basis. The full Hamiltonian matrix can be denoted as $\mH=\l \Bas| H | \Bas\r$ if we interpret $\l \Bas|$ as a column vector of bra-functions. 
The full wave function is $\Psi=\Bas \Coef$.

Here one can clearly see that the basis Eq.~(\ref{eq:recursiveBasis}) reduces to a tensor product, only if the bases at all subnodes of
$J_l$ are equal, $\Bas\upp{J_lj}\equiv\Bas\upp{J_l0},\forall j$:
\begin{equation}\label{eq:tensorBasis}
\Bas^{J_{l}}= 
\left( b^{J_l}_{0}\Bas\upp{J_{l}0},\,b^{J_l}_{1} \Bas\upp{J_{l}0},\ldots,\, b^{J_l}_{K_{J_l}-1} \Bas\upp{J_{l}0}\right)
=\bas\upp{J_l}\otimes \Bas\upp{J_{l}0}.
\end{equation}

For each pair $I_l,J_l$ of index subtrees we define a sub-block $\mH^{I_jJ_j}$ of $\mH=\mH^{I_0,J_0}$,
where the $1\times1$ blocks $\mH^{I_{L}J_{L}}$ are the matrix elements.
The recursive structure of the coefficients induces a recursive block structure of the matrix as
\begin{equation}\label{eq:recursiveOp}
\mH^{I_l,J_l}=
\bma
\mH^{I_l0,J_l0}&\mH^{I_l0,J_l1}&\cdots&\mH^{I_l0,J_lK_{J_l}-1}\\
\mH^{I_l1,J_l0}&\mH^{I_l1,J_l1}&\cdots&\mH^{I_l1,J_lK_{J_l}-1}\\
&\vdots& & \vdots &\\
\mH^{I_lK_{I_l}-1,J_l0}&\mH^{I_lK_{I_l}-1,J_l1}&\cdots&\\
\ema.
\end{equation}
This structure can be exploited for construction of the operator matrices and also for simple representation of 
block-sparsity, e.g. in presence of selection rules. 
One can phrase this recursively as ``an operator matrix is a matrix of operator matrices''. 

If the operator $H$ is a tensor product 
\begin{eqnarray}
H&=&h_0\otimes h_1\otimes\ldots\otimes h_{L-1}=:
h_0\otimes\ldots\otimes h_l\otimes H_{l+1}\nonumber\\
&&H_{l}:=h_l\otimes H_{l+1},\quad l=\range{L}
\end{eqnarray}
the operator matrix for $I_{l},J_{l}$ on level $l$ 
can be assembled from all blocks at the next-lower level $(I_{l+1},J_{l+1})=(I_{l}i,J_{l}j)$ as
\begin{equation}\label{eq:recursionMatrix}
\left[\l\Bas^{I_l}|H_l|\Bas^{J_l}\r\right]_{ij}
=\left[\mH^{I_l,J_l}\right]_{ij}=\l b\upp{I_l}_{i}|h_l|b\upp{J_l}_{j}\r  \mH^{I_{l}i,J_{l}j},
\end{equation}
where $[\ldots]_{ij}$ designates the $ij$-block of $\mH^{I_l,J_l}$ and the indices $i$ and $j$ range from 0 to
$K_{I_l}-1$ and $K_{J_l}-1$, respectively. In practice, the matrix is not usually constructed explicitly.

The tensor-product form of $H_l$ implies a tensor-product form of
$\mH^{I_lJ_l}=\l \Bas^{I_l}|H_l|\Bas^{J_l}\r$, only if also $\Bas^{I_l}$ and $\Bas^{J_l}$ are strict tensor products as in
Eq.~(\ref{eq:tensorBasis}), in which case Eq.~(\ref{eq:recursionMatrix}) reduces to 
\begin{equation}\label{eq:tensorMatrix}
\mH^{I_lJ_l}
=\l \bas\upp{I_l}|h_l|\bas\upp{J_l}\r \otimes \mH^{I_{l}0,J_{l}0}.
\end{equation}
Yet, also when the matrix is not a tensor product, the recursive structure Eq.~(\ref{eq:recursionMatrix}) largely preserves 
the computational advantages of tensor products in terms of the floating point count and, to a lesser degree, data compression.
A typical algorithm for matrix-vector multiplication is discussed in Sec.~\ref{sec:quadratures}.

Many operators in physics can be written as short sums of tensor products and allow efficient and transparent 
computation of the matrices using this scheme. 
In some cases it is advantageous to exploit the recursive structure for applying the
operator matrices to coefficient vectors, as for the radial kinetic energy in two-particle problems. 
If the matrix is very block-sparse, as e.g. in case of dipole selection rules, direct block-wise application performs better.
The choice between these options is made automatically in tRecX based on the actual operator, using non-rigorous 
heuristics.
When operators do not have tensor-product structure, such as electron repulsion $|\vr_1-\vr_2|\inv1$ in the 
Helium atom, the recursive scheme is still used in tRecX for bookkeeping and for ensuring a uniform construction
of operator matrices.

For numerical efficiency, operators are
not usually expanded to the lowest level, but rather recursion is terminated at a ``floor''level $F\leq L$
such that the smallest operator block has typical sizes  of $10\times10 \sim 400\times 400$, depending on the actual problem and
choice of the discretization.
An example for large floor blocks is for the Helium atom, Sec.~\ref{sec:exampleHelium6d}. There the floor level is
put to the two-dimensional finite element patches $[r_{n1},r_{n+1,1}]\times[r_{m1},r_{m+1,1}]$, with a typical number
of $K=20$ functions for each radial coordinate.  Operators on the floor level are usually not represented by full matrices.
In the Helium Hamiltonian Eq.~(\ref{eq:hamHelium}) the first two terms are trivial tensor products. With basis size $K$ on 
both coordinates $r_1,r_2$, the operations count of matrix-vector multiplies is $\order{K^3}$ when one exploits the tensor-product form,
rather than $\order{K^4}$ for general full matrix.
Such structures are automatically recognized by tRecX and implemented using derived classes of an abstract
base class \lcode{OperatorFloor}. Electron repulsion on this lowest level requires application of matrices that are diagonal
for each multipole term with matrix-vector operations count $\order{K^2}$. 
As mentioned above and discussed in Ref.~\cite{zielinski16:doubleionization}, this is not exact, but turns 
out to be an excellent approximation. The high computational cost of electron repulsion 
arises not from the radial part, but from the significant fill-in of the block sparse matrix  by
widely coupling the angular momenta of the two individual electrons. This can be controlled to some extent by truncating 
the multipole expansion at less than maximal order (input \lcode{OperatorFloorEE:lambdaMax}).

The recursive scheme for operators and coefficients translates into simple and transparent algorithms for matrix setup and
matrix-vector operations, which are implemented in the C++  \lcode{class OperatorTree} discussed in Sec.~\ref{sec:classOperator}.

\subsection{Quadratures}
\label{sec:quadratures}
The code makes extensive use of numerical quadrature. This is so, by definition, for FE-DVR basis functions, 
but we apply it throughout: integrals involving trigonometric functions, the associated Legendre 
functions $\aleg{m}{l}$ or the general multi-dimensional basis functions of Sec.~\ref{sec:offCenter} are usually 
all computed by quadratures. Wherever possible, exact quadrature is used. Apart from providing a uniform 
and comparatively error-safe computational scheme in the code, 
exact quadratures are often numerically more stable the evaluation of 
complicated algebraic expressions for analytic integrals.

The tree-structure of the expansion provides for efficient conversion to and from product grids that tRecX uses in 
multi-dimensional quadratures. 
The wave-function value at one point  $Q_{A_0}=(q^0_{\al_0},\ldots,q^{L-1}_{\al_{L-1}})$ of an $L$-dimensional product grid
is
\begin{align}
\Psi&(q^0_{\al_0},\ldots,q^{L-1}_{\al_{L-1}})=\nonumber\\
&\sum_{j_0=0}^{K_{J_0}} b^{J_0}_{j_0}(q^0_{\al_0})
\sum_{j_1=0}^{K_{J_1}} b^{J_1}_{j_1}(q^1_{\al_1})
\cdots
\sum_{j_{L-1}=0}^{K_{J_{L-1}}} b^{J_{L-1}}_{j_{L-1}}(q^{L-1}_{\al_{L-1}})
C\upp{\rangesub{j}{L}}
\end{align}
We abbreviate the matrix of basis function values at the grid points as $b\upp{J_l}_{j_l}(q^{l}_{\al_{l}})=:\mvals{l}$ and 
introduce the intermediate vectors 
\begin{eqnarray}
\mgrid{l}=\Grid\upp{J_{l}}_{\al_lA_{l+1}}
&=&\sum_{j_l=0}^{K_{J_{l}}} \mvals{l}\cdots\sum_{j_{L-1}=0}^{K_{J_{L-1}}} \mvals{L-1}
C\upp{\rangesub{j}{L}}
\nonumber\\
&&
=\sum_{j_l=0}^{K_{J_l}} \mvals{l}\Grid_{A_{l+1}}^{J_{l}j_l}\quad\forall\al_l
\label{eq:toGrid}
\end{eqnarray}
with $A_{l}=(\rangeSub{\al}{{l}}{L-1})$ and the previously defined $J_l=(\rangeSub{j}{0}{l-1})$.
The last equality defines a recursion  starting from coefficients $C\upp{J_L}=\mgrid{L}$ 
and ending at the vector $\mgrid{0}$ of the values of $\Psi$ at all grid points. 

The analogous recursion can be set up for the back-transformation from grid to basis functions.
With quadrature weights $w^l_{\al_l}, \al_l=0,\ldots,S_l-1$ at the grid points $q^l_{\al_l}$ 
one computes the overlap matrices for coordinate $q^l$  at the nodes $J_l$
\[
s^{J_l}_{ij}=\sum_{\al_l}  (b^{J_l})\adj_{i\al_l}w^l_{\al_l}b^{J_l}_{\al_lj}
\]
and from that the factor matrices for back-transformation 
\[
d^{J_l}_{j_l\al_l}=\left[(s^{J_l})\inv1 b^{J_l\dagger}\right]_{j_l\al_l}w^l_{\al_l}.
\] 
On complex-scaled coordinates, the adjoint $b^{J_l\dagger}$
must be replaced by the transpose $b^{J_lT}$, see Sec.~\ref{sec:complexSymmetric}.
The recursion for back-transformation  from $\Grid=\mgrid{0}$ to $C\upp{J_L}=\mgrid{L}$ proceeds by
\begin{equation}\label{eq:fromGrid}
\mgrid{l+1}=\Grid\upp{J_lj_l}_{A_{l+1}}=\sum_{\al_l=0}^{K^{J_l}_l-1} d^{J_l}_{j_l\al_l}\Grid_{\al_lA_{l+1}}^{J_{l}}
\quad\forall j_l
\end{equation}
Both recursions (\ref{eq:toGrid}) and (\ref{eq:fromGrid}) share the same structure and are implemented in a 
\lcode{class OperatorMap}, Sec.~\ref{sec:otherOperators}. 

The recursive algorithm for the transformation to a product grid is very similar to the algorithm for applying a 
tensor product of operators to a vector and it has the same favorable operations count. 
For an ideal quadrature grid with $S_l=K_{J_l}$, the transformation maintains size $\text{len}(\Coef)=\text{len}(\Grid)$
and the operations count for the transformation $\Coef\to\Grid$ is
\begin{equation}
\left(\sum_{l=0}^{L-1} K_{J_l}\right)\times\text{len}(\Coef).
\end{equation}
The computational gain increases exponentially with dimension $L$ comparing to direct application of 
the full transformation matrix $(\prod_{l=0}^{L-1} K_{J_l})\times\text{len}(\Coef)$.
In practice the number of quadrature points often exceeds the number of basis functions, $S_l>K_{j_l}$,
with a corresponding increase of operations count. One prominent example are the associated Legendre functions 
$\aleg{m}{l}(\eta),l\leq L$ where we use a Legendre quadrature grid $\eta_k,k=0,\ldots,L$ which is shared among all $m$ and 
is exact for all overlaps, but inflates the vector length from $L-|m|+1$ to $L+1$. These are more points than, e.g., 
in a Lebedev quadrature grid \cite{lebedev1976}, 
but the product structure is maintained and with it the efficient algorithm for transformation to the grid.

For simplicity we have treated the case where all coordinates are transformed to a grid, but obviously transformations can be
limited to a given subset of the coordinates, as needed. In tRecX, the creation of product grids and 
transformations to and from them are handled by a specialized class 
\lcode{DiscretizationGrid}, see Sec.~\ref{sec:mainClasses}.

\subsection{Adaptive features}
\label{sec:adaptive}

In problems that are strongly driven by the external field, time step size and required basis size can 
change significantly as the system evolves. Step sizes decrease near field peaks and 
increase near field nodes. By default, the code automatically controls the size of the time
steps based on a standard single-to-double step estimate, which has an overhead slightly above 50\%. We have decided to use this 
simple but universal control algorithm, which only requires a well-defined consistency order of the underlying 
time-stepper, in order to maintain flexibility in choosing the time-stepper. In strongly driven systems, 
gain by adaptive step size can be up to a factor of 2 compared to a step fixed at the maximal stable size. 
The maybe more important advantage of step size control in tRecX is that well-defined accuracies are achieved without 
the need of careful time-step adjustment.
At the end of time propagation average step size and its variance are printed, based on which one can fix the step 
size once a system's behavior in a given parameter range and discretization is known.

A typical phenomenon of strong-field physics is a large increase in angular momenta as the field ramps up. After the 
end of the pulse, those angular momentum components gradually decay and the operator does not need to be applied to them.
Also, in absence of the pulse the interaction part of the operator is zero. These developments are monitored in the code and 
operators are only applied in regions where there is non-negligible contribution to the time-evolution. Control is achieved
by estimating the contribution to the derivative vector based on the norm of the floor operator block $||\mH\upp{I_FJ_F}||$,
which is precomputed at setup, and a norm of the \rhs vector $\Coef_{J_F}$. As the vector norm needs to be evaluated
at every time-step, we use the simple estimate $||\Coef_{J_F}||_a: = \max_i[|\Re(C_i)|+|\Im(C_i)|]$. If the contribution to 
the total vector norm is below a threshold $||\mH\upp{I_FJ_F}|| ||\Coef_{J_F}||_a\leq\ep_{th}$ application of the block is skipped.
The procedure requires some care with choosing $\ep_{th}$, but can speed up computations by factors $\lesssim 2$ without loss
of accuracy. Application is demonstrated in \lcode{tutorial/13}. 
The code will print some advice when $\ep_{th}$ may have been chosen too large or too small, 
but at present heuristics for the choice of $\ep_{th}$ is incomplete. By default $\ep_{th}=0$, i.e. blocks are only skipped when 
the operator block or the vector become exactly zero, which happens, for example, after the end of a laser pulse
with strictly finite duration. 

\subsection{Control of stiffness}
\label{sec:stiffness}
For time-propagation at present only explicit methods are used, whose efficiency notoriously deteriorates
as the norm of the operator matrix increases. The main origin of large norm in 
Schr\"odinger-like problems is the Laplacian,
whose matrix norm grows as $p_{\max{}}^2\sim \de x\inv2$, where $p_{\max{}}$ and  $\de x$ are the characteristic 
scales of maximal momentum and spatial resolution, respectively. Usually one does not manage 
to restrict the momenta in the discretization to the physically relevant level and spurious, very high eigenvalues
appear that can dramatically slow down explicit time-steps to the level of numerical breakdown of the propagation. This stiffness
problem can be fixed, if one manages to remove spurious eigenvalues from the operators. In tRecX,
high-lying eigenvalues of the field-free Hamiltonian are suppressed by spectral projections. In the simplest form one 
replaces the full Hamiltonian matrix with a projected one
\begin{equation}
\mH(t)\to (\one-\mP) \mH(t) (\one-\mP),\qquad \mP=\sum_i |i\r \l i|, 
\end{equation}
where the $|i\r$ are orthonormal eigenvectors for large eigenvalues of the field-free Hamiltonian.

With more challenging Hamiltonians like for the
Helium atom or molecular systems, the full field-free Hamiltonian has many high-lying spurious states, they are expensive 
to compute, and application of the projection becomes costly. In such cases one can use for the $|i\r$ eigenvectors of
a different operator, for example the Laplacian. Eigenvectors of the Laplacian are sparse due to 
rotational symmetry and in case of multi-particle systems they can be given as tensor products of single-electron vectors. 
This renders calculation of the eigenvectors as well as application of the projection computationally cheap. 

The cutoff energy for removal 
of high-lying states is characteristically set around 100 \au. This is far larger than the actual energy scale
of typically $\lesssim 10\,\au$ However, choosing the threshold that low would compromise the results and raises the cost
of applying the projection to the point where no compute time is gained, in spite of the fact that step-size increases 
inversely proportional to the cutoff energy.
The energy cutoff is first introduced in \lcode{tutorial/07}. Examples for using the Laplacian
instead of the full Hamiltonian for projecting are in \lcode{tutorial/21} and \lcode{23}. 

Comparing to an outright spectral representation of the operators, removing a small number of outlier eigenvalues from a 
local representation maintains all sparsity deriving from locality of operators represented in a local basis. The cost of 
removal remains low because the number of removed eigenvalues is small compared to the basis size and the vectors may have 
tensor product form, as for the Laplacian of the He atom.

\subsection{Parallelization}
\label{sec:parallel}
tRecX is parallelized using MPI, but it will also compile without MPI, if no MPI library is detected by Cmake. 
The code is aware of hardware hierarchy in that it can distinguish between ``compute nodes'' assumed connected 
through switches, ``boards'' connected by a bus, and ``CPUs'' assumed to have fast shared memory access. 
This hierarchy, although present in the code, is not at present exploited by the distribution algorithm.
For local operators, communication between non-overlapping elements of the FE-DVR  is low. Operator locality 
between elements is detected during setup and taken into account by the default distribution algorithms 
for the respective coordinate systems.

The finest MPI grains are the \lcode{OperatorFloor} blocks $\mH^{I_FJ_F}$. These operate between 
 subsections of the coefficient vectors $\Coef_{I_F}$ and  $\Coef_{J_F}$ with typical dimensions 10 to 400.
The operator blocks can be distributed arbitrarily across all MPI nodes, but 
communication overhead must be taken into consideration. The corresponding \lcode{class OperatorFloor} 
has a member \lcode{cost()} that
determines the CPU load for its application by self-measurement during setup. A heuristic
algorithm uses these numbers to create a load-balanced distribution of the operator. For containing communication cost,
care is taken to arrange blocks into groups that share either $I_F$ or $J_F$.
At least one of the respective sections of coefficient vectors $\Coef_{J_F}$ or $\Coef_{J_F}$ 
reside on the same parallel process, which then ``owns'' the corresponding $I_F$ or $J_F$.

Actual communication cost is not measured by the code. Rather, it assumes there is a sorting of the  $\Coef_{I_F}$ 
such that compunction is dominantly short range, as e.g. sorting by increasing angular momenta in case of dipole interactions.
Then neighboring $\Coef_{I_F}$'s are preferably assigned to the same thread. The default for this sorting is 
the sequence how the \lcode{Axis:name} are input. For some coordinate systems this is overridden by internal 
defaults and the user can in turn can override by the input \lcode{Parallel:sort}. 
The sorting actually used is shown in the output.

\subsubsection{Scaling}
\label{sec:scaling}
Problems that can be solved with tRecX vary widely in structure and also in the methods employed. Scaling
behavior strongly depends on these choices. Memory is, in general, not a limiting factor for tRecX calculations.
Parallelization strategy focuses on large problems where run times in sequential mode would be days or weeks, while little
effort has been made to boost parallelization for small problems with runtimes on the scale of minutes. Into  the latter category 
fall many problems in tRecX that would be on time scales of hours with more traditional approaches.
These gains in program efficiency are through complex features such as exploiting tensor products and block-sparsity, 
by stiffness control, the use of high order methods, or the tSurff box-size reduction. All these features, while at times dramatically 
reducing compute times and problem sizes, tend to lead to coarser graining and enhanced communication, which necessarily 
deteriorates scalability. Specifically the haCC method is inherently non-local with large communication and therefore mostly restricted 
to shared memory use. 

\begin{figure}\label{fig:scaling}
\includegraphics[width=0.47\textwidth]{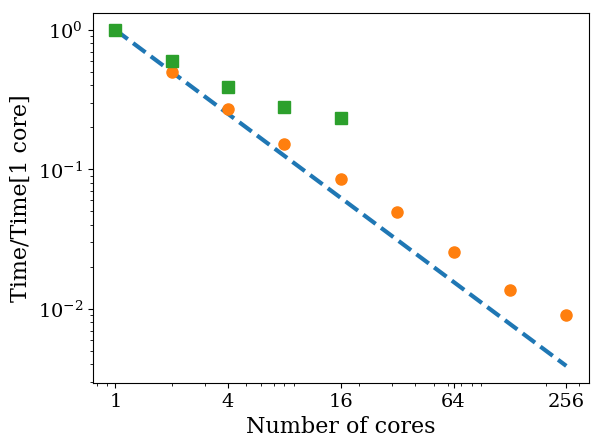}
\caption{Scaling of tRecX for a double-ionization calculation of the Helium atom at total basis size of $3\times10^5$ (dots) and 
single-electron problem discussed in Sec.~\ref{sec:exampleSingle} at basis size $1460$ (squares).}
\end{figure}

Most problems treated with tRecX are best solved on small parallel machines in the range from 4 to 64 cores.
Only large problems such as the double-ionization of the Helium atom can profit from more extensive parallelization. 
Fig.~\ref{fig:scaling} shows the scaling behavior for fixed-size problems (``strong scaling''). The two examples 
are hydrogen in an IR field discussed in Sec.~\ref{sec:exampleSingle} 
and a Helium atom computation with 20 $l$-functions, $m=-1,0,1$ and 91 radial functions for 
each electron, resulting in total basis size of $3\times 10^5$. 
Computations were performed at the LMU Theory machine KCS hosted at the 
Leibnitz Rechenzentrum (LRZ), which consists of compute nodes connected by infiniband and 
dual boards with 2$\times$16 cores on each node. Parallelization gains can be seen up to 256 cores. 
Scaling remains away from linear and as always in this situation one has to weigh time gains for individual computations 
against overall throughput for multiple runs.

\section{Main classes}
\label{sec:mainClasses}
Here we discuss the classes that form the functional and conceptional backbone of tRecX. A complete listing of all 
classes is provided by the code's Doxygen documentation. In general, many classes in
the code have a \lcode{.write()} member for dumping to file and a matching \lcode{.read()} or constructor for recovery 
from file. Mostly for debugging purposes, there is usually a \lcode{.str()} member that returns a human-readable 
\lcode{string}. Also, for critical classes, there are \lcode{test()} functions that provide
cross-checks and usage examples. A key role is played by the abstract template class \lcode{Tree}.

\subsection{Index class}

The C++ \lcode{class Index} represents the complete recursive basis tree defined through (\ref{eq:recursiveBasis}). The class and its member data are declared as
\begin{lstlisting}
class Index: public Tree<Index> {
    mutable uint32_t _size; 
    uint16_t _indexBas;
    uint8_t _indexAx;
    char _indexKind;
    ....
}
\end{lstlisting}
These  member data refer to the given node and are the only index-specific data of the tree. The complete 
tree-structure, such as tree iterators, pruning, transposition, and other tree transformations are 
implemented in the  \lcode{template class  Tree}, Sec.~\ref{sec:classTree}, which is used for all tree classes of tRecX.
The \lcode{Index} data is squeezed into 8 bytes in an attempt to minimize storage, 
as \lcode{Index} trees can become very large. This limits the number of different single-level basis sets $\bas$ that are pointed to 
by \lcode{_indexBas} to $2^{15}$. In practice, also in very large computations only a few tens of different bases appear.  
Whenever a tensor product basis is used, the same basis re-appears at many nodes and has the same \lcode{_indexBas}, 
as for example the product bases for the 
$r_1,r_2$-discretization. The range of the axis pointer \lcode{_indexAx} is $2^7$, which is sufficient as the number of axes is intimately 
related to the dimension of the problem and hardly ever exceeds 10. \lcode{_size} gives the length of $\Coef_J$ at the node.
This information is redundant, but is cached here for fast access, and similarly \lcode{_indexKind} is cached information about the
node's function and position within the tree.

The \lcode{Index} class, as one of the code's oldest classes, is burdened by legacy code. In order to disentangle the 
current from legacy code, primary construction is through an auxiliary derived class \lcode{IndexNew} which takes an 
\lcode{AxisTree} as its input. \lcode{AxisTree}, in turn, reflects the definitions read from input.  The standard \lcode{AxisTree}  
is trivial with a single branch per node, equivalent to a vector. Only when hybrid discretizations are used, 
as for the molecular problem (Sec.~\ref{sec:exampleMolecule}) and for off-centers bases (Sec.~\ref{sec:offCenter}), 
the tree becomes non-trivial.

An \lcode{Index} contains the complete information about the basis $\Bas$, Eq.~(\ref{eq:recursiveBasis}). 
It also has a member function \lcode{overlap()} that returns a pointer to 
$\l\Bas\upp{J_l}|\Bas\upp{J_l}\r$ as long as this is 
a meaningful entity for a single $\Bas\upp{J_l}$, i.e. when the subspace on level $l$ does not have overlap with 
any other subspace on the same level $\l\Bas\upp{I_l}|\Bas\upp{J_l}\r=0$ for $I_l\neq J_l$.

\subsubsection{Special \lcode{Index} constructors}
\label{sec:indexSpecial}
\begin{figure}[h]
\includegraphics[height=6cm]{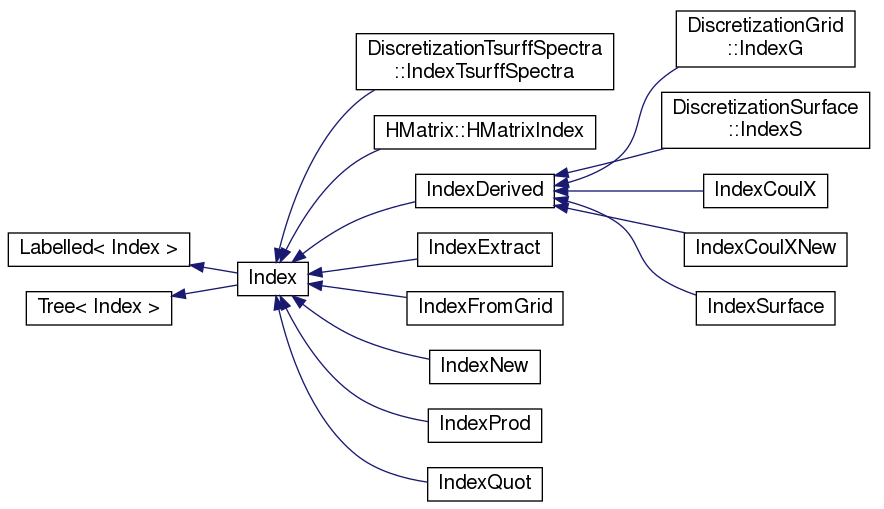}
\caption{\label{fig:Index} Hierarchy of \lcode{Index} classes, sec~\ref{sec:indexSpecial}.}
\end{figure}

There is a number of specialized constructors for indices, see Fig.~\ref{fig:Index}.  For disentangling from the 
legacy code, these are usually given as the constructor of a derived class. One useful constructor is \lcode{IndexG} 
for transforming from basis functions to representation by grid values. 
The grid can be equidistant, useful for plotting, or a quadrature grid, which is convenient for various forms
of basis transformations and quadratures. The tree-structure of the basis allows to perform these transformations
computationally efficiently, cf. Sec.~\ref{sec:quadratures}. The \lcode{IndexS} represents value and radial derivative at the tSurff surface.
It is constructed from a standard \lcode{Index} by specifying the coordinate axis name and the surface radius $R_s$. 
\lcode{IndexProd} constructs a new index tree as the tensor product of two \lcode{Index} trees.
\lcode{IndexQuot} forms the ``quotient'' of an \lcode{Index full} by a ``denominator'' \lcode{Index den} 
by eliminating from \lcode{full} all basis levels that appear in \lcode{den}, ensuring consistency of the result. This allows to extract, 
e.g., a single-electron factor from a two-electron basis. 
Maps to these derived indices are generated automatically by a class \lcode{OperatorMap} (see below) 
and are not invertible in general

\subsubsection{\lcode{Discretization} classes}
Various classes derived form \lcode{class Discretization} are wrappers around specific indices and serve as
an interface for \lcode{Index} construction. Examples are \lcode{DiscretizationGrid} (Sec.~\ref{sec:quadratures}) and 
\lcode{DiscretizationTsurffSpectra} (Sec.~\ref{sec:otherOperators}). 

An important derived class is \lcode{DiscretizationSpectral}, which
constructs all or a selected part of the eigenvalues and eigenvectors of any diagonalizable operator matrix $\mA$ and presents them in the
form of a diagonal operator $\md_A$ (\lcode{class OperatorDiagonal}). Further it contains 
transformations $\mU$ and $\mV$ (\lcode{class OperatorMap}) from and to the original basis, respectively. Note that in general $\mU\neq\mV\adj$ as the original basis as a rule is not orthonormal and also $\mA$ may not be hermitian. 
Block-diagonal structure of the original operator is recognized and translated into block-diagonal transformation. Arbitrary 
functions of the eigenvalues can be formed using \lcode{OperatorDiagonal}. 
E.g. one can form $\exp(-it\mA)=\mU\exp(-i\md_A)\mV$ for time-integration of small problems, or, similarly, to implement rotations 
in a spherical harmonic basis. The principal use in tRecX is in stiffness control, Sec.~\ref{sec:stiffness}.

For operators of the special form $\oH=\oH_I \otimes \one + \one \otimes \oH_J$ the class \lcode{DiscretizationSpectralProduct}
constructs a spectral representation taking full advantage of the fact that there is an eigenbasis of $\oH$ in the form 
of a tensor product of eigenbases of $\oH_I$ and $\oH_J$. Transformations to and from that spectral representation have 
tensor product form. The class can also be used when the basis $\Bas$ is not tensor product, but is related to a tensor product
$\Bas\upp{I}\otimes\Bas\upp{J}$ by a constraint as in Sec.~\ref{sec:basisConstraint}.

\subsection{The template class Tree}
\label{sec:classTree}

All trees in the code are derived from \lcode{class Tree} by the "curiously recursive template pattern" exemplified
in \lcode{class Index:public Tree<Index>}. We list a few key features of this class, but refer to the documented code
for the complete definition and functionality.

\lcode{Tree} has the private data
\begin{lstlisting}
template <typename T> class Tree {
    const T* _parent;
    vector<T*> * _child;
    ...
\end{lstlisting}
that point to a node's parent and all its children and are accessed through member functions \lcode{parent()} and \lcode{child(int j)}, respectively. 
Iterators along various paths through the tree are provided. For legacy reasons these are not realized in the standard C++
iterator syntax, but rather by member functions returning pointers to the incremented node. The most important iterators are \lcode{descend()}
for descending from a node to its left-most branch and \lcode{nodeRight(Origin)} the next node to the right within the subtree
originating at node \lcode{Origin}. 

Nodes without branches are called leafs. A standard sorting of leafs is by their position along the lower edge of the tree. 
The functions \lcode{firstLeaf()} and \lcode{nextLeaf()} return leftmost leaf descending from a given node and the iterator through the leafs. Note that
in general \lcode{nextLeaf()} is not equivalent to \lcode{nodeRight()} as a tree's lower edge does not need to remain at
the same level depth, as in the example of Fig.~\ref{fig:indexTree}
The index of a node is returned by \lcode{vector<int> index()}. For class \lcode{Index}, this 
is exactly the tuple $J_l$ defined in Sec.~\ref{sec:discretization}. 

Functions to add and remove branches include \lcode{childAdd(T* C)} and \lcode{childPop()}. For re-sorting trees there
is a \lcode{permute(...)}, which takes a permutation of the tree levels as its argument and returns at tree with 
the levels permuted. A typical case would be the transposition of tensor indices. With non-tensor objects, 
as e.g.\ in Sec.~\ref{sec:basisConstraint}, it may not be
possible to interchange certain indices in an unambiguous way and an exception will be raised upon the attempt.

Finally, trees can also be realized as ``views'', which do not actually own copies of their data, but rather point do 
data of another tree. This is particularly useful for re-arranging tree data into a new tree by permuting indices without 
actually moving the data.

\subsection{Operators classes}
\label{sec:classOperator}
All operator classes are derived from an abstract base class with the following key data and member function:
\begin{lstlisting}
class OperatorAbstract {
    const Index *iIndex, *jIndex; 
    void apply(complex<double> A,const Coefficients&X
                complex<double>B,Coefficients&Y)const=0;
\end{lstlisting}
It symbolizes a map $\vX\to\vY=Op(\vX)$. 
The class containing the coefficients is a tree
\lcode{class Coefficients: public Tree<Coefficients>} which is usually constructed from
an \lcode{Index* idx} as \lcode{Coefficients X(idx)}. It mirrors the tree structure of \lcode{idx} and, at
each of its nodes $J_l$, it points to the data of $\vX_{J_l}$.  
Derived classes must implement the virtual abstract function \lcode{apply(...)} for the map $\vY\leftarrow A\vX + B\vY$.
On this abstract level, there are no particular assumptions other than that the operator maps from a linear space into a 
linear space. The two spaces do not need be equal or subspaces of the same space, 
the map itself does not need to be linear.

\begin{figure}[h]
\includegraphics[width=\textwidth]{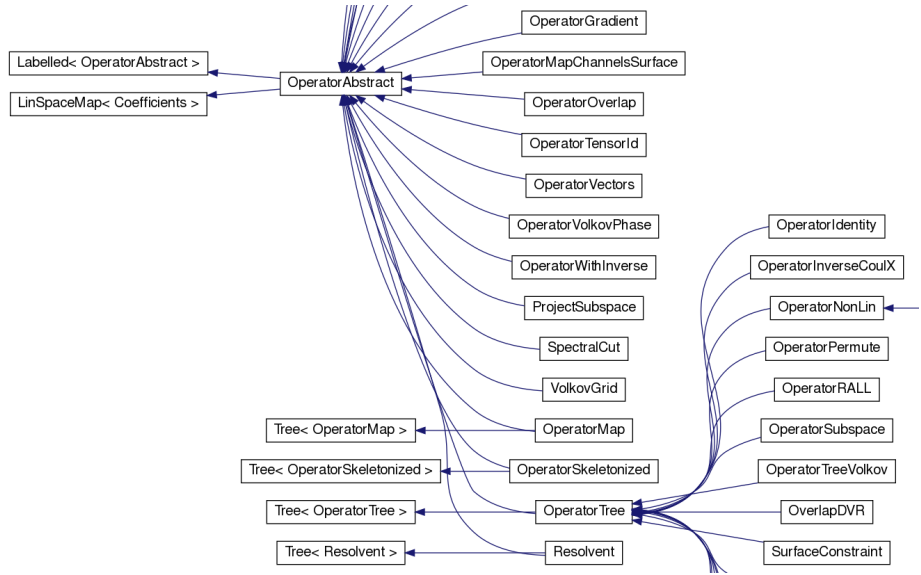}
\caption{\label{fig:OperatorAbstract} Hierarchy of operators in tRecX (incomplete). An \lcode{OperatorAbstract} is an instance of a map between linear spaces \lcode{LinSpaceMap}, specifically between \lcode{Coefficients}. Implementations  
\lcode{OperatorTree}, \lcode{OperatorMap}, and \lcode{Resolvent} are briefly described in the text.}
\end{figure}
A large number of diverse operators are derived from \lcode{OperatorAbstract}, part of who are shown in the 
Doxygen-generated class hierarchy in Fig.~\ref{fig:OperatorAbstract}. Particularly important is
\begin{lstlisting}
class OperatorTree: public Tree<OperatorTree>, 
                    public OperatorAbstract {
protected:
    OperatorFloor * oFloor;
    ...
\end{lstlisting}
which implements the hierarchy of block-matrices (\ref{eq:recursiveOp}). The \lcode{oFloor} pointer is only 
non-null at the leafs of the operator tree. The \lcode{class OperatorFloor} implements 
all forms of maps in a numerically efficient way, for example, multiplication of a vector by 
a full or diagonal matrix, multiplication by a tensor product of small matrices, but also more complicated
maps as, for example in the electron-electron interaction for a given multipole-contribution. Again, the map may be
also non-linear, as in a Gross-Pitaevskii operator. These various forms are realized as derived classes of the 
abstract base class \lcode{OperatorFloor}. 

\subsubsection{Construction and optimization of an \lcode{OperatorTree}}
The primary constructor of \lcode{OperatorTree} takes an
operator definition string as in the examples of Sec.~\ref{sec:applications} that matches \lcode{Index} and recursively
sets up the full operator. As a rule, no complete matrix is constructed. Mostly, the tree contains only the non-zero
\lcode{OperatorFloor}'s. If tensor product structure is detected in the operator, 
it is exploited if found to be numerically advantageous by some (approximate) internal algorithm. 
When multiple terms contribute to the same \lcode{OperatorFloor} these are summed into a single \lcode{OperatorFloor}
where this is possible and numerically profitable. 

\subsubsection{Further important operator classes}
\label{sec:otherOperators}
From the whole list of operators we further single out the following classes for their more general relevance:
\paragraph{\lcode{OperatorInverse}}
Calculates the inverses of overlap matrices using Woodbury-like methods
consisting of a cheap direct inverse with some low rank update for completing the exact inverse.
\paragraph{\lcode{MapGauge}}
Implements general Gauge transformations.
\paragraph{\lcode{OperatorMap}}
Given two \lcode{Index} objects for discretizations $\Bas$ and $\Bas'$ this is the map $\Bas\to\Bas'$, where this is logically
possible and meaningful. Typical examples are maps to and from grids, Sec.~\ref{sec:quadratures}. 
Another application is in \lcode{class DiscretizationTsurffSpectra} for the transformation from surface values 
to a grid of momentum points, where the momentum spectra are accumulated. Such transformations are not necessarily lossless.
\paragraph{\lcode{Resolvent}}
Given $\mH$ and an overlap $\mS$ as \lcode{OperatorAbstract}'s, this class 
constructs the resolvent operator $(\mH-z\mS)\inv1$ with a complex $z$. 
At present the implementation is through \lcode{Eigen}'s sparse LU-decomposition
and is limited by basis sizes. For banded matrices, like in the Floquet example discussed above, \lcode{Resolvent} can be
constructed for very large dimensions.

\subsection{Recursive algorithms}

Recursive structures provide for tRecX's flexibility, but in addition they generate compact and comparatively 
transparent code. The basic pattern is:
\begin{lstlisting}
function(tree):
    if isLeaf: specific action on leaf
    else: for c in children: function(c)
\end{lstlisting}

As examples, we discuss the \lcode{apply} member function of the class \lcode{OperatorTree} and 
an \lcode{OperatorTree} constructor. 
For \lcode{apply}, which implements $\vY\leftarrow b \vY + a \mO[\vX]$, the simplified pseudo-code is
\begin{lstlisting}
1: O.apply(a,X,b,Y):
2:   Y<-b*Y
3:   if isLeaf: 
4:     OperatorFloor.apply(a,X,1,Y)
5:   else:
6:     for cO in O.children:
0:       cO.apply(a,X.child(cO.jdx),1,Y.child(cO.idx))
\end{lstlisting}
Here an \lcode{OperatorFloor} class implements the specialized action in a efficient way, typically through LAPACK or Eigen. 
The code plays out its efficiency for block-sparse matrices, where zero-blocks never appear in the loop, line 6. 
The price to pay is that one needs to locate the block operator's left- and right hand indices in the coefficient vectors $\vX$ and $\vY$,
here symbolical written as \lcode{X.child(cO.jdx)} and \lcode{Y.child(cO.idx)}, respectively. 
If one ensures that \lcode{OperatorFloor.apply} is a sufficiently coarse-grain operation, say multiplication by a 20$\times$20 matrix, 
the overhead from the recursion remains small. Clearly, for full matrices or matrices with very regular structure 
such as band-matrices, the algorithm is at a disadvantage. Where such performance losses are identified, \lcode{OperatorTree} should be replaced by a more specialized class derived from \lcode{OperatorAbstract}.

After all setup is done  the \lcode{OperatorTree} a ``flattened'' view of the tree is created for use in propagation.
This is a vector of pointers to the leafs of the \lcode{OperatorTree}, which are automatically distributed for parallelization (see sec.~\ref{sec:parallel}).
In the process, direct pointers from the operator indices to the respective sections of the $\vX$ and $\vY$ vectors are set up, 
eliminating all overhead from that place.

A second example of recursion in tRecX is the pseudo-code of a basic 
\lcode{OperatorTree} constructor, for the case of a strict tensor product
\lcode{opDef="0.5<def0><def1>...<defL>"}
\begin{lstlisting}
OperatorTree(opDef,iIndex,jIndex,mult): 
  # e.g., a,f,r=0.5,"<def0>","<def1>...<defL>"
  a,f,r=getFactors(opDef) 
  
  # mat(i,j)=<b[i]|op[f]|b[j]>
  Matrix mat=getMatrix(f,iIndex.basis,jIndex.basis)  
  
  if r=="":  
    mat*=a*mult
    oFloor=OperatorFloor(mat)
  else where mat(i,j)!=0:
    mult*=a*mat(i,j)
    childAdd(OperatorTree(r,
             iIndex.child(i),jIndex.child(j),mult)
\end{lstlisting}
The \lcode{opDef} strings are split into the scalar prefactor \lcode{a}, the first
tensor factor \lcode{f}, and the remainder \lcode{r} by \lcode{getFactors}. 
This is mainly located in class \lcode{OperatorDefinition}, with a few additional classes due to legacy code.
Then \lcode{getMatrix} interprets the tensor factor string \lcode{s=<defN>} and constructs the corresponding factor matrix.
If one has arrived at the last factor, the remainder \lcode{r} becomes empty. The matrix \lcode{mat} is multiplied by scalar
factors \lcode{mult} and \lcode{a} and its matrix-vector application is implemented depending on its structure, e.g.,
for full, diagonal or banded matrices. If the remainder is not empty one advances to the next tensor factor. In this simple example, 
tensor structure is multiplied out rather than preserved.

The actual tRecX code is more complex, admitting for tensor products, the sum of terms, and handling of special operators such as
\lcode{[[eeInt6DHelium]]} in Sec.~\ref{sec:coulombRepulsion}.
Also syntax and consistency of the defining string \lcode{opDef} with the actual left and right indices \lcode{iIndex,jIndex} 
are checked throughout and errors throw exceptions. At the end of construction
the \lcode{OperatorTree} is post-processed where multiple operators for the same index pair are fused into single blocks
and zero blocks that may have appeared after summation are eliminated.

\subsection{Basis sets}
All bases are derived from a class \lcode{BasisAbstract} with the pure virtual function \lcode{size()} giving the number 
of functions in the basis. The word ``basis'' is used in a general way for any set of defining properties 
for the discrete representation on a given $q^l$. This includes a discrete set of functions, but also grids, 
or an orthonormal set of unit vectors in a discrete space. Bases need not be orthonormal, although this is ensured 
wherever it is possible and meaningful.

\subsubsection{\lcode{BasisIntegrable}}
\lcode{BasisIntegrable} is an abstract class is for single-variable basis functions that can be integrated over:
\begin{lstlisting}
class BasisIntegrable: public BasisAbstract {
protected:
    double _lowBound,_upBound;
public:
    virtual void valDer(
        const vector<complex<double> > &X,
        vector<complex<double>> &V,
        vector<complex<double>> &D,...) const=0;
    virtual void quadRule(
        int N, vector<double> &QuadX, 
        vector<double> &QuadW) const=0;
    virtual unsigned int order() const=0; 
    ...
\end{lstlisting}
The functions are supported on the interval [\lcode{_lowBound},\lcode{_upBound}], which may also be infinite.
The pure virtual function \lcode{valDer(...)} must be implemented to return the value and first derivative matrices 
$V_{ij}=b_j(x_i)$ and $D_{ij}=b'_j(x_i)$. Any \lcode{BasisIntegrable} must provide $N$-point quadrature rules
$q_k,w_k$ in \lcode{QuadX,QuadW} through \lcode{quadRule(...)}. Finally, there is the concept ``order'' of a \lcode{BasisIntegrable}. This can be understood as the minimal number of quadrature points needed for the correct
evaluation of overlap matrix elements. For example, in 
a DVR basis with Dirichlet boundary conditions an the lower boundary, the first Lagrange polynomial is omitted, leading 
to \lcode{size()=order()-1}.

A simple example of a \lcode{BasisIntegrable} that is only used for debugging purposes are the monomials
$\{1,x,x^2,\ldots\}$:
\begin{lstlisting}
class BasisMonomial:public BasisIntegrable{
  int _order;
public:
  BasisMonomial(int Order, double Low, double Up)
      :BasisIntegrable(Low,Up),_order(Order){}
  unsigned int size() const{return _order;}
  
  void quadRule(...) const{
    ...shift-and-scale Legendre quadrature...
  }
  
  void valDer(const vector<complex<double>> &X,
              vector<complex<double>> &V,
              vector<complex<double>> &D,...
              ) const {
    V.assign(X.size(),1.);
    D.assign(X.size(),0.);
    for(int k=1;k<size();k++){
      for(int i=0;i<X.size();i++){
        V.push_back(V[X.size()*(k-1)+i]*X[i]);
        D.push_back(D[X.size()*(k-1)+i]*X[i]
                   +V[size()*(k-1)+i]       );
      }}}
      
  unsigned int order() const{return _order;}
};
\end{lstlisting}
Note the use of ``automatic differentiation'' for the evaluation of the derivatives. This transparent and 
efficient approach to determining derivatives is used throughout tRecX.

\subsubsection{\lcode{BasisDVR}}
\label{sec:basisDVR}
An important \lcode{BasisIntegrable} implementation is \lcode{BasisDVR}, where the most important data members are
\begin{lstlisting}
class BasisDVR: public BasisIntegrable{
    vector<double> _dvrX,_dvrW;
    int _nBeg; 
    int _size;
     ...
\end{lstlisting}
\lcode{_dvrX} and  \lcode{_dvrW} are the nodes and weights for quadrature
rule. The rule is Lobatto for finite intervals and Radau for semi-infinite intervals. 
There are at most \lcode{_dvrX.size()} different Lagrange polynomials, for which values 
and derivatives can be evaluated anywhere within the basis' interval. Dirichlet boundary conditions
are determined through \lcode{_nBeg} and \lcode{_size}. \lcode{_nBeg}=0 means the Lagrange polynomial 
for \lcode{_dvrX[0]=_lowBound} is included, and =1, where that polynomial is omitted. Similarly, 
\lcode{_nBeg+_size=_dvrX.size()-1} means the Lagrange polynomial at \lcode{_dvrX.back()=_upBound} is omitted
for Dirichlet condition at that point. The values are set upon construction.

\subsubsection{\lcode{BasisGrid}}
\lcode{BasisGrid} is not a \lcode{BasisIntegrable}, rather it derives directly from \lcode{BasisAbstract}:
\begin{lstlisting}
class BasisGrid: public BasisAbstract {
    vector<double> _mesh;
    unsigned int size() const {return _mesh.size();}
    ...
\end{lstlisting}
The only class-specific member data is \lcode{_mesh}, which holds the grid points.
Values of a \lcode{BasisGrid} are only defined at the grid points, but a member function for 
Newton-interpolation between these points is provided.

The class is mostly for transforming \lcode{BasisIntegrable}'s to grids. Assume a
wave function $|\Psi\r=|\Bas\r \Coef$ is given in terms of an \lcode{Index cIdx} containing \lcode{BasisIntegrable}'s.
A new \lcode{IndexG gridIdx(cIdx)} is created where the \lcode{BasisIntegrable}'s are replaced by the desired
\lcode{BasisGrid}'s. In the process an \lcode{OperatorMap mapFrom} 
is automatically created which transforms $\Psi$ to its representation on the multi-dimensional grid. Assuming
\lcode{Coefficients X(cIdx)} contains the $\Coef$, then
\begin{lstlisting}
mapFromParent()->apply(1,X,0,Y)
\end{lstlisting}
fills \lcode{Coefficients Y(gridIdx)} with $\Psi(\vx_i)$ at the multi-dimensional grid points $\vx_i$.
The assignment between values $\Psi(\vx_i)$ and $\vx_i$ is given through the structure information contained
in \lcode{gridIdx}.

A class \lcode{BasisGridQuad} is derived from \lcode{BasisGrid}, with an additional member
\lcode{vector<double> _weights} for integration weights at the \lcode{_mesh}. This allows
a lossless transformation between of \lcode{BasisIntegrable} to a Gauss quadrature 
grid that is exact for the basis. This procedure is used on several occasions, e.g., for the efficient 
multiplication by the Volkov phases on a grid of $\vk$-values (see  Sec.~\ref{sec:tsurff}), when the spectral amplitudes 
are given in terms of spherical harmonics.

\subsubsection{\lcode{BasisVector}}
\lcode{BasisVector} is a simple and useful class for discrete coordinate indices,
where only $l$ matters and the value of $q^l$ has no significance. It is fully defined by its size
\begin{lstlisting}
class BasisVector : public BasisAbstract{
    unsigned int _size;
    unsigned int size() const{return _size;}
    ...
\end{lstlisting}
This is used, for example to label the Floquet blocks in sec.~\ref{sec:floquet}.

\subsubsection{\lcode{BasisSub}}
A subset of a given \lcode{BasisAbstract} is selected by \lcode{BasisSub} 
\begin{lstlisting}
class BasisSub: public BasisAbstract{
    vector<int> _subset;
    const BasisAbstract* _bas;
    ...
\end{lstlisting}
where the \lcode{vector<int>_subset} lists function numbers from \lcode{_bas} to be included with \lcode{BasisSub}. 
This is used when imposing basis constraints or
in general when pruning branches from an \lcode{Index}.

\subsubsection{Multi-dimensional basis functions --- \lcode{BasisNdim}}

As illustrated in Sec.~\ref{sec:exampleMolecule} and further discussed in Sec.~\ref{sec:discretization}, formal coordinates $q^l$ may
also be multi-dimensional. Functions with higher-dimensional arguments appear as orbitals but
also as intermediate objects when mixing coordinates systems, 
for example for a multi-center expansion. The class is more complex than the examples given so far. The general strategy is
to store the values and partial derivatives of all functions at a suitable quadrature grid. This may require substantial 
memory, but we have not exhausted standard size storage of a few GB in applications so far. The quadrature grid may refer
to a different coordinate system \lcode{_quadCoor} than the basis's coordinate system \lcode{_ndimCoor}. 
From this follows the class signature:
\begin{lstlisting}
class BasisNdim : public BasisAbstract{
    string _ndimCoor; 
    string _quadCoor; 
    vector<vector<double>> _quadGrid;
    vector<double> _quadWeig;
    vector<vector<vector<complex<double>>>> _valDer; 
    ...
\end{lstlisting}
Matrix elements can be computed for operators given in terms of standard strings, where the transformation between
different coordinate systems is done automatically adhering to the philosophy of ``automatic differentiation''. For further
details we refer to the in-line documentation of the code.

\subsection{Input, units conversion, and algebraic expressions}

\change{}{
An attempt is made to make input human-readable, error-safe, and self-explanatory. Rather than listing available inputs
in some separate manual, the code itself enforces input documentation and input sanity checks. Erroneous input 
triggers error messages showing line number in the input file and valid options, emits a warning about suspicious
input or throws a run-time errors when inconsistent input is detected. A dynamically generated list of possible input is displayed, 
when the code is run without parameters. In general, plausible guesses for the input will be accepted or trigger information 
on valid alternatives. More details are shown in the following.}

\subsubsection{General input format --- class \lcode{ReadInput}}

All user input is controlled by a \lcode{class ReadInput} with a prescribed designation of any input item in the format
\lcode{Category:}~\lcode{name} as illustrated in the examples of Sec.~\ref{sec:applications}. An overloaded \lcode{read(...)} method requires to supply a
default value or to state explicitly that there cannot be a default and brief documentation for every input item. 
Inputs can be of all standard types, which also includes 
\lcode{vector}'s. Input is usually read from file but can be overruled by command line flags of the format \lcode{-Category:name=value} by 
default or abbreviated flags, that can be specified in \lcode{read(...)}. 
In the input file, several \lcode{name}'s can follow the \lcode{Category} specifier, the sequence of the names is arbitrary.
Also, the same category can appear in repeated lines, such as in Sec.~\ref{sec:exampleMolecule} where we have
\lcode{Operator:hamiltonian} and \lcode{Operator:interaction}. We follow the convention of having \lcode{Category}'s 
start with upper case and \lcode{name}'s with lower case letters.

There is a simple syntax to restrict admissible input values.
A member function \lcode{ReadInput::finish()} checks all inputs from file and from the command line for correct \lcode{Category} and \lcode{name} and will stop if a given pair \lcode{Category:name} in the file does not actually appear in the code, reducing the 
likelihood of misprint errors. In addition, \lcode{finish()} a list of all admissible inputs in a file \lcode{tRecX.doc}, 
which also explains the input as documented in \lcode{read(...)} and the default input values. The contents of this
file is shown as a help when running \lcode{tRecX} without any input. 

The above is meant to illustrate the general strategy for enforcing documentation and  enhancing usability and error safety. 
Full features can be found in the inline-documentation and are illustrated by a usage example in the 
\lcode{test()} member function.

Another feature for productivity and error safety is the possibility to freely choose input units and 
to use algebraic expressions as inputs. Default are \au unless the input name specifies a different 
unit. In the example of section 
\ref{sec:exampleMolecule}
\begin{lstlisting}
Laser:shape,I(W/cm2),FWHM,lambda(nm), phiCEO,polarAngle
       cos2,  1e10, 4 OptCyc, 800,      pi/2,   45
       cos4,  1e11, 3 OptCyc, 800/13,   0,      45
       cos4,  1e11, 3 OptCyc, 800/15,   0,      45
\end{lstlisting}
the intensity is expected with the strong-field convention as $W/cm^2$. The units in 
brackets at \lcode{I(W/cm2)} form a functional part of the \lcode{Category:}\lcode{name}. The \lcode{name}'s input units can be overruled by
specifying e.g. \lcode{1e-2 au} instead. That value will be converted by \lcode{read(...)} to $W/cm^2$,
$10\inv2 \au =3.52\ldots\times10\inv{14}W/cm^2$ with full available precision. Another example is with 
\lcode{lambda(nm)}, where, e.g. one could equivalently use the input string \lcode{800e-9 m}.

\subsubsection{Class \lcode{Units}}

Unit conversions are performed by a \lcode{class Units} which at present recognizes atomic units \lcode{au}, cgs \lcode{ESU},
and \lcode{SI} units plus a few units that customarily used in strong field physics such as $W/cm^2$, $nm$ and Rydberg energy \lcode{Ry}. 
The duration of an optical cycle \lcode{OptCyc} is computed from the wave-length of the field component in the first line after \lcode{Laser}: 
the listing above produces
1~\lcode{OptCyc}~$=2\pi (800\,nm)/c$ (converted to \au).

\subsubsection{Class \lcode{Algebra}}
\label{sec:algebra}
Input values can be specified as algebraic expressions of constants, as in \lcode{800/13} or \lcode{pi/2}. The strings
are interpreted by the same \lcode{class Algebra} that is used for the definition of operators. It can do standard complex algebra, where complex numbers are specified as in \lcode{2+i*3.1415}. It recognizes a few constants such as \lcode{pi} and \lcode{hbar} 
($\hbar$ in SI units). Further constants can be added from the input, as documented in the command line help. 

\change{}{
When used for constructing functions of a single coordinate, the character \lcode{Q} represents the coordinate in expressions such 
as \lcode{pow[2](cos(Q/2))}, which on a \lcode{Phi}-axis would evaluate to $\cos^2(\phi/2)$. The most frequent mathematical 
functions are available, see the \lcode{tutorial}s for examples. When attempting to input a malformed algebra, 
a diagnostic of the error will be displayed which also includes the full list of presently implemented functions.
}
\subsection{\lcode{TimePropagator} and \lcode{TimePropagatorOutput} classes}

Time propagation is controlled through a wrapper class \lcode{TimePropagator}. It takes start and end times, accuracy 
or step size and output intervals as its main control parameters. For solving the ordinary differential equation in time it
needs a class of abstract type \lcode{ODEstep}. A range of those steppers have been implemented including a general (explicit)
Runge-Kutta, a specialized classical 4-stage Runge-Kutta, and Arnoldi solver, and several experimental solvers. At present, 
only the classical Runge-Kutta is used, as it was found to be the overall most efficient across the large variety of problems 
treated with tRecX. The notorious stiffness problem of explicit methods is controlled by removing 
few extremely high-lying spectral values from the problem, see Sec.~\ref{sec:stiffness}. Although this does deliver a workable and rather efficient solution, we do not consider the development of 
time-steppers as concluded. 

The \lcode{class TimePropagatorOutput} controls which information is output during time-propagation. One category of outputs are
expectation values of operators, by default the overlap $\l \Psi(t)|\Psi(t)\r$ and field-free Hamiltonian 
$\l \Psi(t)|H_0|\Psi(t)\r$, where $H_0$ is the operator specified as \lcode{Operator:hamiltonian}. Further expectation
values can be defined at the input, for example the dipole values in various gauges. Another category
are \lcode{Coefficients} for  $\oA|\Psi(t)\r$.  
In this way the values and derivatives at the tSurff-radius $R_c$ are written to disc, but $\oA$ can also be user-defined.
More transformations can be easily added by editing \lcode{main_trecx.cpp}.

\subsubsection{\lcode{Plot}}
One can plot densities of the kind $|\Psi(\vq,t)|^2$ or more generally $\overline{\Psi}(\vq,t)[\oA\Psi](\vq,t)$ 
with a user-defined operator $\oA$. This is handled by 
\lcode{class Plot}, which is constructed from input as, for example,
\begin{lstlisting}
Plot: axis,points,lowerBound,upperBound
      Rn,101,0.,20.
      Eta,31,-1.,1.
\end{lstlisting}
where the density is plotted \wrt the discretization's axes \lcode{Rn} and \lcode{Eta} the two-dimensional region 
$[0,20]\times[-1,1]$ with $101\times 31$ equidistant grid points. Coordinates not listed are assumed to be integrated or 
summed over. Output will be in ASCII format and readable, e.g., by Gnuplot, but also by tRecX's \lcode{plot.py} script.
The order of inputs lines in \lcode{Plot} determines the sorting of the density values, 
such that the first \lcode{axis}, \lcode{Rn} in the example, runs fastest. For higher-dimensional plots the further 
dimensions will appear as additional columns in the two-dimensional output file.
Explanation of the input for plots can be found in \lcode{tRecX.doc}, for the full features of the class we refer to the Doxygen 
and inline documentation of the code.

\subsection{Python scripts}
\label{sec:scripts}

There is a limited number of convenience python scripts in the \lcode{SCRIPTS} subdirectory.
These have mostly grown out of practice and certainly do not comply with good programming requirements. 
Yet, given their proven usefulness in practice, we include them with the distribution.

For submission to compute queues one can adjust \lcode{submit_tRecX.py}, which is currently set up for SLURM and 
should be adaptable to similar queuing systems with little effort. It ensures generation of properly named run-directories
before starting the actual tRecX code. By this one can submit multiple jobs without the need to manually ensure proper 
run-directory numbering. It also creates a short submit name for the job for display by the SLURM queue overview.

Virtually all ASCII files that appear in the run directory can be plotted using \lcode{plot.py}. It produces one- and two-dimensional graphs 
from selected columns of a file, compares multiple runs, can annotate curves with the actual parameters used in the run etc. 
Brief instructions and a full list of command line flags are displayed by running  \lcode{plot.py} without any arguments.

Running multiple calculations with varying parameters, either for ensuring convergence or for analyzing a physical phenomenon is a 
frequent mode of using tRecX. The script \lcode{lRuns.py} lists all or a selected subset of runs showing 
basic information such as status of the computation, run time, wave function norm, and energy. In addition, the user can select 
any set of input parameters for display. Usage instructions are shown when running \lcode{lRuns.py} on the command line without any parameters.

\hide{
\section{Advanced and experimental features}
\tikzstyle{decision} = [diamond, draw, fill=blue!20, 
    text width=4.5em, text badly centered, node distance=3cm, inner sep=0pt]
\tikzstyle{block} = [rectangle, draw, fill=blue!20, 
    text width=5em, text centered, rounded corners, minimum height=4em]
\tikzstyle{line} = [draw, -latex']
\tikzstyle{cloud} = [draw, ellipse,fill=red!20, node distance=3cm,
    minimum height=2em]
\begin{tikzpicture}[node distance = 2cm, auto]
    \node [block] (init) {initialize model};
    \node [cloud, left of=init] (expert) {expert};
    \node [cloud, right of=init] (system) {system};
    \node [block, below of=init] (identify) {identify candidate models};
    \node [block, below of=identify] (evaluate) {evaluate candidate models};
    \node [block, left of=evaluate, node distance=3cm] (update) {update model};
    \node [decision, below of=evaluate] (decide) {is best candidate better?};
    \node [block, below of=decide, node distance=3cm] (stop) {stop};
    \path [line] (init) -- (identify);
    \path [line] (identify) -- (evaluate);
    \path [line] (evaluate) -- (decide);
    \path [line] (decide) -| node [near start] {yes} (update);
    \path [line] (update) |- (identify);
    \path [line] (decide) -- node {no}(stop);
    \path [line,dashed] (expert) -- (init);
    \path [line,dashed] (system) -- (init);
    \path [line,dashed] (system) |- (evaluate);
\end{tikzpicture}

\section{Libraries}
}

\section{Conclusions}

The purpose of tRecX is three-fold: applications, training and education, and community development.

The code produces accurate solutions for TDSEs that appear in ultrafast and strong field physics. 
In the present public version a wide range of standard problems such as high harmonic generation, fully 
differential spectra for single ionization, 
Floquet and various model systems can be solved by adapting the given tutorial inputs. 
Also, with the use of significant computer resources, fully differential double 
emission spectra can be computed. With tSurff as one of its key methods, computer resource consumption remains low, 
on the scale of a few minutes for single-electron calculation of standard tasks, and within the range of the feasible for long-wavelength
double emission. Forthcoming releases will include haCC, 
which integrates Gaussian-based quantum chemical wave functions with the discretizations discussed here.
This allows to compute emission from multi-electron systems.

A designated part of the tRecX development is to ensure user experience that is acceptable to a somewhat wider range of specialist users, including
experimentalists who want to generate standard results or study simple models as well as theorists with more complex demands. We 
consider error safe and intuitive input, extensive consistency checks, and structurally enforced documentation as essential for achieving
that goal. 

Finally, on the developer level, the systematic C++ object orientation has allowed development and maintenance of the code
by a very small group. The full research code is also used in training and education on the undergraduate and 
graduate level. In course of such projects, attention to understandable and consistent code structure it taught and enforced. 
Student projects have 
non-trivially contributed to the code in specialized applications, such as the use of parabolic coordinates, Coulomb scattering, and
double- and triple breakup (not included in the public release yet).

\change{}{
For standard use, tRecX in its present form will be made available at the ``AMP gateway'', a collaborative effort 
for low-threshold use of atomic physics codes \cite{AMPgateway}. At present, a preliminary 
installation is available a that site.
}

The experience with student projects shows that substantial structural contributions from a community
are possible without endangering code integrity or maintainability. 
Possible first such projects would likely be collaborative, but
also unsupervised extensions may well be feasible. A formal invitation for contributions is extended here.
\section*{Acknowledgment}
Key initial contributions to the code were made by Vinay Pramod Majety and Alejandro Zielinski, with 
further contributions by, in alphabetic order, Christoph Berger, Jonas Bucher, Florian Egli, Jacob Liss, Mattia Lupetti, 
J\"orn St\"ohler, Jonathan Rohland, Andreas Swoboda, Hakon Volkmann, Markus and Michael Weinmueller,  and Jinzhen Zhu.
Funding was provided by the DFG excellence cluster EXC 158 ``Munich Center for Advanced Photonics'' (MAP), the Austrian Science
Foundation project ViCoM (F41) and the DFG priority program 1840 (QUTIF).

\bibliographystyle{unsrt}
\bibliography{../../bibliography/photonics_theory}

\end{document}

%% file: my_commands.tex
\usepackage{amsmath}
\usepackage{amssymb}
\usepackage{makeidx}
\usepackage{graphicx}
\usepackage{color}

\newcommand{\wrt}{w.r.t.\ }
\newcommand{\lhs}{lhs.\ }
\newcommand{\rhs}{rhs.\ }

\newcommand{\hide}[1]{}

\newcommand{\inv}[1]{ ^{-#1}}

\newcommand{\order}[1]{\mathcal{O}(#1)}

\newcommand{\myvec}[1]{\vec{#1}}
\newcommand{\vr}{{\myvec{r}}}
\newcommand{\vx}{{\myvec{x}}}

\newcommand{\vk}{{\myvec{k}}}

\newcommand{\vq}{{\myvec{q}}}

\newcommand{\vX}{{\myvec{X}}}
\newcommand{\vY}{{\myvec{Y}}}

\newcommand{\val}{{\myvec{\alpha}}}

\newcommand{\vb}{{\myvec{b}}}
\newcommand{\vA}{{\myvec{A}}}

\newcommand{\vC}{{\myvec{C}}}

\newcommand{\vPhi}{{\myvec{\Phi}}}

\newcommand{\vunit}[1]{\hat{#1}}

\newcommand{\uep}{\vunit{\epsilon}}

\newcommand{\myoper}[1]{\mathbf{#1}}
\newcommand{\oA}{\myoper{A}}

\newcommand{\oH}{\myoper{H}}

\newcommand{\mymat}[1]{{\widehat{#1}}}

\newcommand{\mA}{\mymat{A}}

\newcommand{\mH}{\mymat{H}}

\newcommand{\md}{\mymat{d}}
\newcommand{\mO}{\mymat{O}}
\newcommand{\mP}{\mymat{P}}

\newcommand{\mS}{\mymat{S}}

\newcommand{\mU}{\mymat{U}}
\newcommand{\mV}{\mymat{V}}

\newcommand{\range}[1]{0,1,\ldots,#1-1}

\newcommand{\rangesub}[2]{#1_0,#1_1,\ldots,#1_{#2-1}}
\newcommand{\rangeSub}[3]{#1_#2,\ldots,#1_{#3}}
\newcommand{\rangesup}[2]{#1^0,#1^1,\ldots,#1^{#2-1}}
\newcommand{\rangeSup}[3]{#1^#2,\ldots,#1^{#3}}

\newcommand{\RR}{\mathbb{R}}
 
\newcommand{\one}{\mathbf{1}} 

\newcommand{\adj}{^\dagger}

\renewcommand{\r}{\rangle}
\renewcommand{\l}{\langle}

\newcommand{\ddt}{\frac{d}{dt}}

\newcommand{\ddx}{\partial_x}

\newcommand{\ddy}{\partial_y}
\newcommand{\ddz}{\partial_z}

\newcommand{\ddr}{\partial_r}

\newcommand{\om}{\omega}

\newcommand{\ep}{\epsilon}
\newcommand{\al}{\alpha}

\newcommand{\Ga}{\Gamma}
\newcommand{\ga}{\gamma}
\newcommand{\de}{\delta}

\newcommand{\De}{\Delta}
\renewcommand{\th}{\theta}

\newcommand{\lcase}{\left\{\begin{array}{ll}}
\newcommand{\rcase}{\end{array}\right.}
\renewcommand{\bar}{\begin{array}{ll}}
\newcommand{\ear}{\end{array}}
\newcommand{\bal}{\begin{align}}
\newcommand{\eal}{\end{align}}
\newcommand{\bma}{\begin{pmatrix}}
\newcommand{\ema}{\end{pmatrix}}
\newcommand{\beq}{\begin{equation}}
\newcommand{\eeq}{\end{equation}}
\newcommand{\bel}[1]{\begin{equation}\label{eq:#1}}
\newcommand{\eel}{\end{equation}}
\newcommand{\bea}{\begin{eqnarray}}
\newcommand{\eea}{\end{eqnarray}}
\newcommand{\beaNN}{\begin{eqnarray*}}
\newcommand{\eeaNN}{\end{eqnarray*}}

\newcommand{\vEf}{\vec{\mathcal{E}}}

\newcommand{\vna}{\vec{\nabla}}

\newcommand{\pde}{\partial}
